\documentclass[aps,prd,twocolumn,amssymb,superscriptaddress,floatfix,nofootinbib]{revtex4}

\usepackage{graphicx}
\usepackage{dcolumn}
\usepackage{bm}
\usepackage{epsfig}
\usepackage{float} 

\providecommand{\KPpnn}{\mbox{${K}^+\to\pi^+\nu\bar\nu$}}
\providecommand{\KZpnn}{\mbox{${K}^0_{\rm L}\to\pi^0\nu\bar\nu$}}

\providecommand{\KPtwo}{\mbox{$K_{\pi2}$}}

\providecommand{\KPpnothing}{\mbox{$K^+\to\pi^+\!\!+\!{nothing}$}}

\providecommand{\pnnone}{\mbox{$\pi\nu\bar{\nu}(1)$}}
\providecommand{\pnntwo}{\mbox{$\pi\nu\bar{\nu}(2)$}}
\providecommand{\oneortwo}{\mbox{$\pi\nu\bar{\nu}(1\!\!+\!\!2)$}}

\providecommand{\KIN}{\mbox{${\rm KIN}_{\rm\bf T}$}}
\providecommand{\KINBAR}{\mbox{${\rm KIN}_{\rm\bf R}$}}
\providecommand{\KINSTD}{\mbox{${\rm KIN}_{\rm\bf  S}$}}
\providecommand{\TD}{\mbox{${\rm TD}_{\rm\bf T}$}}
\providecommand{\TDBAR}{\mbox{${\rm TD}_{\rm\bf R}$}}
\providecommand{\PV}{\mbox{${\rm PV}_{\rm\bf T}$}}
\providecommand{\PVBAR}{\mbox{${\rm PV}_{\rm\bf R}$}}
\providecommand{\DC}{\mbox{${\rm DC}_{\rm\bf T}$}}
\providecommand{\DCBAR}{\mbox{${\rm DC}_{\rm\bf R}$}}

\providecommand{\KINSTD}{\mbox{${\rm KIN}_{\rm\bf  S}$}}
\providecommand{\PVSTD}{\mbox{${\rm PV}_{\rm\bf  S}$}}
\providecommand{\DCSTD}{\mbox{${\rm DC}_{\rm\bf  S}$}}
\providecommand{\TDSTD}{\mbox{${\rm TD}_{\rm\bf  S}$}}

\providecommand{\BRSM}{\mbox{$(0.85\pm0.07)\times 10^{-10}$}}  
\providecommand{\BRthis}{\mbox{$(7.89^{+9.26}_{-5.10})\times10^{-10}$}} 
\providecommand{\BRall}{\mbox{$(1.73^{+1.15}_{-1.05})\times 10^{-10}$}} 
\providecommand{\UCLall}{\mbox{$3.35\times10^{-10}$}}   
\providecommand{\UCLgn}{\mbox{$14.6\times10^{-10}$}} 

\providecommand{\BRscalar}{\mbox{$(9.9^{+8.5}_{-4.2})\times10^{-10}$}}
\providecommand{\BRtensor}{\mbox{$(4.9^{+3.9}_{-2.4})\times10^{-10}$}}
\providecommand{\UCLscalar}{\mbox{$21\times 10^{-10}$}} 
\providecommand{\UCLtensor}{\mbox{$10\times 10^{-10}$}} 

\providecommand{\TOTbkgd}{\mbox{$0.927\pm0.168{}^{+0.320}_{-0.237}$}}
\providecommand{\TOTbkgdTRIM}{\mbox{$0.93\pm0.17({\rm stat.}){}^{+0.32}_{-0.24}({\rm syst.})$}} 
\providecommand{\KBLIVE}{\mbox{$1.71\times 10^{12}$}} 

\newcommand{\subsubsubsection}[1]{ {\noindent\normalsize\textbf{\textsl{ #1}} } }

\begin{document}

\preprint{BNL-81786-2008-JA} 
\preprint{FERMILAB-PUB-09-007-CD-T} 
\preprint{KEK/2008-44} 
\preprint{TRIUMF/TRI-PP-08-26} 
\preprint{UHEP-EX-08-004} 

\title{\boldmath Study of the decay \KPpnn\  in the momentum region 
  $140<P_\pi<199\ {\rm MeV}/c$}


\author{A.V.~Artamonov}\affiliation{Institute for High Energy Physics, Protvino, Moscow Region, 142 280, Russia}
\author{B.~Bassalleck}\affiliation{Department of Physics and Astronomy, University of New Mexico, Albuquerque, NM 87131}
\author{B.~Bhuyan}\altaffiliation{Now at Department of Physics, Indian Institute of Technology Guwahati, Guwahati, Assam, 781 039, India.}\affiliation{Brookhaven National Laboratory, Upton, NY 11973}
\author{E.W.~Blackmore}\affiliation{TRIUMF, 4004 Wesbrook Mall, Vancouver, British Columbia, Canada V6T 2A3}
\author{D.A.~Bryman} \affiliation{Department of Physics and Astronomy, University of British Columbia, Vancouver, British Columbia, Canada V6T 1Z1}
\author{S.~Chen} \affiliation{Department of Engineering Physics, Tsinghua University, Beijing 100084, China} \affiliation{TRIUMF, 4004 Wesbrook Mall, Vancouver, British Columbia, Canada V6T 2A3} 
\author{I-H.~Chiang} \affiliation{Brookhaven National Laboratory, Upton, NY 11973}
\author{I.-A.~Christidi}\altaffiliation{Now at Physics Department, Aristotle University of Thessaloniki, Thessaloniki 54124, Greece} \affiliation{Department of Physics and Astronomy, Stony Brook University, Stony Brook, NY 11794}
\author{P.S.~Cooper}\affiliation{Fermi National Accelerator Laboratory, Batavia, IL 60510}
\author{M.V.~Diwan} \affiliation{Brookhaven National Laboratory, Upton, NY 11973}
\author{J.S.~Frank} \affiliation{Brookhaven National Laboratory, Upton, NY 11973}
\author{T.~Fujiwara}\affiliation{Department of Physics, Kyoto University, Sakyo-ku, Kyoto 606-8502, Japan}
\author{J.~Hu} \affiliation{TRIUMF, 4004 Wesbrook Mall, Vancouver, British Columbia, Canada V6T 2A3}
\author{J.~Ives} \affiliation{Department of Physics and Astronomy, University of British Columbia, Vancouver, British Columbia, Canada V6T 1Z1} 
\author{D.E.~Jaffe} \affiliation{Brookhaven National Laboratory, Upton, NY 11973}
\author{S.~Kabe} \affiliation{High Energy Accelerator Research Organization (KEK), Oho, Tsukuba, Ibaraki 305-0801, Japan}
\author{S.H.~Kettell} \affiliation{Brookhaven National Laboratory, Upton, NY 11973}
\author{M.M.~Khabibullin}\affiliation{Institute for Nuclear Research RAS, 60 October Revolution Prospect 7a, 117312 Moscow, Russia}
\author{A.N.~Khotjantsev}\affiliation{Institute for Nuclear Research RAS, 60 October Revolution Prospect 7a, 117312 Moscow, Russia}
\author{P.~Kitching} \affiliation{Centre for Subatomic Research, University of Alberta, Edmonton, Canada T6G 2N5}
\author{M.~Kobayashi} \affiliation{High Energy Accelerator Research Organization (KEK), Oho, Tsukuba, Ibaraki 305-0801, Japan}
\author{T.K.~Komatsubara} \affiliation{High Energy Accelerator Research Organization (KEK), Oho, Tsukuba, Ibaraki 305-0801, Japan}
\author{A.~Konaka} \affiliation{TRIUMF, 4004 Wesbrook Mall, Vancouver, British Columbia, Canada V6T 2A3}
\author{A.P.~Kozhevnikov}\affiliation{Institute for High Energy Physics, Protvino, Moscow Region, 142 280, Russia}
\author{Yu.G.~Kudenko}\affiliation{Institute for Nuclear Research RAS, 60 October Revolution Prospect 7a, 117312 Moscow, Russia} 
\author{A.~Kushnirenko} \altaffiliation{Now at Institute for High Energy Physics, Protvino, Moscow Region, 142 280, Russia.}  \affiliation{Fermi National Accelerator Laboratory, Batavia, IL 60510} 
\author{L.G.~Landsberg}\altaffiliation{Deceased.}\affiliation{Institute for High Energy Physics, Protvino, Moscow Region, 142 280, Russia}
\author{B.~Lewis}\affiliation{Department of Physics and Astronomy, University of New Mexico, Albuquerque, NM 87131}
\author{K.K.~Li}\affiliation{Brookhaven National Laboratory, Upton, NY 11973}
\author{L.S.~Littenberg} \affiliation{Brookhaven National Laboratory, Upton, NY 11973}
\author{J.A.~Macdonald} \altaffiliation{Deceased.} \affiliation{TRIUMF, 4004 Wesbrook Mall, Vancouver, British Columbia, Canada V6T 2A3}
\author{J.~Mildenberger} \affiliation{TRIUMF, 4004 Wesbrook Mall, Vancouver, British Columbia, Canada V6T 2A3}
\author{O.V.~Mineev}\affiliation{Institute for Nuclear Research RAS, 60 October Revolution Prospect 7a, 117312 Moscow, Russia}
\author{M. Miyajima} \affiliation{Department of Applied Physics, Fukui University, 3-9-1 Bunkyo, Fukui, Fukui 910-8507, Japan}
\author{K.~Mizouchi}\affiliation{Department of Physics, Kyoto University, Sakyo-ku, Kyoto 606-8502, Japan}
\author{V.A.~Mukhin}\affiliation{Institute for High Energy Physics, Protvino, Moscow Region, 142 280, Russia}
\author{N.~Muramatsu}\affiliation{Research Center for Nuclear Physics, Osaka University, 10-1 Mihogaoka, Ibaraki, Osaka 567-0047, Japan}
\author{T.~Nakano}\affiliation{Research Center for Nuclear Physics, Osaka University, 10-1 Mihogaoka, Ibaraki, Osaka 567-0047, Japan}
\author{M.~Nomachi}\affiliation{Laboratory of Nuclear Studies, Osaka University, 1-1 Machikaneyama, Toyonaka, Osaka 560-0043, Japan}
\author{T.~Nomura}\affiliation{Department of Physics, Kyoto University, Sakyo-ku, Kyoto 606-8502, Japan}
\author{T.~Numao} \affiliation{TRIUMF, 4004 Wesbrook Mall, Vancouver, British Columbia, Canada V6T 2A3}
\author{V.F.~Obraztsov}\affiliation{Institute for High Energy Physics, Protvino, Moscow Region, 142 280, Russia}

\author{K.~Omata}\affiliation{High Energy Accelerator Research Organization (KEK), Oho, Tsukuba, Ibaraki 305-0801, Japan}
\author{D.I.~Patalakha}\affiliation{Institute for High Energy Physics, Protvino, Moscow Region, 142 280, Russia}
\author{S.V.~Petrenko}\affiliation{Institute for High Energy Physics, Protvino, Moscow Region, 142 280, Russia}
\author{R.~Poutissou} \affiliation{TRIUMF, 4004 Wesbrook Mall, Vancouver, British Columbia, Canada V6T 2A3}
\author{E.J.~Ramberg}\affiliation{Fermi National Accelerator Laboratory, Batavia, IL 60510}
\author{G.~Redlinger} \affiliation{Brookhaven National Laboratory, Upton, NY 11973}
\author{T.~Sato} \affiliation{High Energy Accelerator Research Organization (KEK), Oho, Tsukuba, Ibaraki 305-0801, Japan}
\author{T.~Sekiguchi}\affiliation{High Energy Accelerator Research Organization (KEK), Oho, Tsukuba, Ibaraki 305-0801, Japan}
\author{T.~Shinkawa} \affiliation{Department of Applied Physics, National Defense Academy, Yokosuka, Kanagawa 239-8686, Japan}
\author{R.C.~Strand} \affiliation{Brookhaven National Laboratory, Upton, NY 11973}
\author{S.~Sugimoto} \affiliation{High Energy Accelerator Research Organization (KEK), Oho, Tsukuba, Ibaraki 305-0801, Japan}
\author{Y.~Tamagawa} \affiliation{Department of Applied Physics, Fukui University, 3-9-1 Bunkyo, Fukui, Fukui 910-8507, Japan}
\author{R.~Tschirhart}\affiliation{Fermi National Accelerator Laboratory, Batavia, IL 60510}
\author{T.~Tsunemi}\altaffiliation{Now at Department of Physics, Kyoto University, Sakyo-ku, Kyoto 606-8502, Japan.}\affiliation{High Energy Accelerator Research Organization (KEK), Oho, Tsukuba, Ibaraki 305-0801, Japan}
\author{D.V.~Vavilov}\affiliation{Institute for High Energy Physics, Protvino, Moscow Region, 142 280, Russia}
\author{B.~Viren}\affiliation{Brookhaven National Laboratory, Upton, NY 11973}
\author{Zhe~Wang} \affiliation{Department of Engineering Physics, Tsinghua University, Beijing 100084, China} \affiliation{Brookhaven National Laboratory, Upton, NY 11973} 
\author{N.V.~Yershov}\affiliation{Institute for Nuclear Research RAS, 60 October Revolution Prospect 7a, 117312 Moscow, Russia}
\author{Y.~Yoshimura} \affiliation{High Energy Accelerator Research Organization (KEK), Oho, Tsukuba, Ibaraki 305-0801, Japan}
\author{T.~Yoshioka}\affiliation{High Energy Accelerator Research Organization (KEK), Oho, Tsukuba, Ibaraki 305-0801, Japan}
\collaboration{E949 Collaboration}\noaffiliation

\date{\today}

\begin{abstract}
Experiment E949 at Brookhaven National Laboratory has observed three new
events consistent with the decay \KPpnn\ in the pion momentum region
$140 < P_\pi < 199\ {\rm MeV}/c$ in an exposure of \KBLIVE\ stopped
kaons with an estimated total background of \TOTbkgdTRIM\ events. 
This brings the total number of observed \KPpnn\ events  to seven. 
Combining this observation with previous results, 
assuming the pion spectrum  predicted by the standard model,  
results in  a branching ratio of ${\cal B}(\KPpnn)=\BRall$.
An interpretation of the results for 
alternative models of the decay {\KPpnothing} is also presented.
\end{abstract}


\maketitle

\section{Introduction}
\label{sec:Introduction}  

This article is a  detailed report of the 
final results from experiment E949 at Brookhaven National Laboratory 
on  the study of \KPpnn\ in the pion momentum 
region $140<P_\pi < 199\ {\rm MeV}/c$~\cite{ref:pnn2PRL}. 
The observation of 
{\KPpnothing}, a charged kaon decay to a single charged pion and no
other observable particles, is evaluated within the
framework of the standard model (SM) and in 
terms of alternative models.

\subsection{Interpretation of the decay \boldmath\KPpnothing}
The only significant SM contribution to 
the experimental signature 
\KPpnothing , where $nothing$ represents experimentally unobservable particles,  
 is \KPpnn\ where $\nu\bar\nu$ is $\nu_e \bar \nu_e$,
$\nu_{\mu} \bar \nu_{\mu}$ or $\nu_{\tau} \bar \nu_{\tau}$ 
as   discussed in
Ref.~\cite{ref:pnn1PRD}. The calculation of the branching ratio 
 has undergone  continual
theoretical refinement and experimental narrowing of the relevant
input parameters since the first modern treatment of this
process~\cite{Gaillard:1974hs,Inami:1980fz}.  
A recent assessment of the prediction
for the branching ratio of this process is 
${\cal B}(\KPpnn)={\BRSM}$~\cite{ref:TH2} where the quoted uncertainty is
dominated by the uncertainty in the 
Cabibbo-Kobayashi-Maskawa quark-mixing matrix elements. 
This assessment  included new small
corrections to the charm quark contributions to the SM branching
ratio.

There have been many alternatives to the SM interpretation 
of \KPpnothing\ signature 
suggested over the years,  including 
the following:
\begin{enumerate}
\item 
New
physical mechanisms contributing to \KPpnn\ with the usual
neutrino-antineutrino pairs.  
Many models incorporating 
new physics would result in a deviation from the
SM prediction for \KPpnn.  A summary of these 
through mid-2007 can be found in \cite{Buras:2004uu}.  
Since that time there have been new
calculations of the branching ratio in the littlest Higgs model with
T-parity~\cite{Blanke:2008ac,Goto:2008fj}, the possible effects on
the branching ratio of a heavy singlet up-quark~\cite{Kopnin:2008ca}
and a reassessment of the
constraints of the Minimal Flavor Violation Model~\cite{Hurth:2008jc}.

\item Cases in which the neutrino flavor
is not conserved.
There are examples stemming from extended
Technicolor~\cite{Appelquist:2004ai}, supersymmetry (SUSY)~\cite{Grossman:2003rw}, and
new effective four-fermion interactions involving
neutrinos~\cite{ref:nonSMnu}.  Like most examples of lepton flavor
violation in kaon decay, these tend to be small, but there are cases
such as some types of R-violating SUSY~\cite{Deandrea:2004ae}, in
which \KPpnn~ gives the limiting constraint on some of the couplings.

\item Reactions in which a single unseen particle recoils
against the $\pi^+$.
These include species of axions~\cite{Hindmarsh:1998ph}, the
familon~\cite{Wilczek:1982rv}, sgoldstinos~\cite{Gorbunov:2000th}, 
a gauge boson corresponding to a new ${\rm U}(1)'$  group~\cite{Aliev:1988we,Pospelov:2008zw},
and various light dark-matter 
candidates ~\cite{Pospelov:2007mp,ref:light_dark_matter,Gunion:2005rw}.
In general these models do not predict branching ratios; rather they  use
\KPpnothing\ results to constrain parameters.

\item Other exotic processes.
These include the effects of ``unparticles'', which 
can change the SM $\pi^+$ energy spectrum as well as the branching
ratio~\cite{Wu:2008xg}.
\end{enumerate}

\subsection{Previous results on \KPpnn\ below the $K_{\pi2}$ peak}
A detailed discussion of the history of measurements of \KPpnn\ was
given in~\cite{ref:pnn1PRD}.  However most of these measurements were
made in the kinematic region in which the $\pi^+$ is more energetic
than the $\pi^+$ from the background reaction $K^+\to\pi^+\pi^0$ ($K_{\pi2}$), 
dubbed the ``{\pnnone}'' region.  
\begin{figure} 
\includegraphics[width=\linewidth]{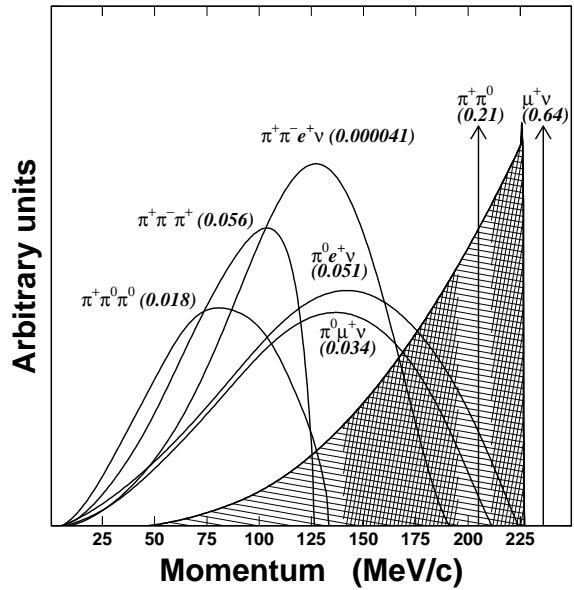}
\caption{\label{fig:seven}Momentum spectra of charged particles 
from $K^+$ decay in the rest frame. 
The values in parentheses represent 
the branching ratios of the decay modes~\cite{ref:PDG}. 
The hatched spectrum represents the $\pi^+$ spectrum from \KPpnn\ decay 
assuming the $V-A$ interaction. The densely hatched regions represent
the \pnnone\ and \pnntwo\ E949 signal regions.}
\end{figure}
By contrast  fewer measurements have been made in the 
``{\pnntwo}'' region in which the $\pi^+$ is less energetic than that from
$K_{\pi2}$ (Figure~\ref{fig:seven}).  As will be discussed below, this region is experimentally 
more challenging than the \pnnone\  region for a stopped-kaon geometry 
principally because 
the $\pi^+$ from $K_{\pi2}$ decay can enter the \pnntwo\ region
if it undergoes a nuclear interaction in the stopping target.

Among the examples of \pnntwo\  measurements was the first attempt to
measure \KPpnn\ in  a heavy liquid bubble chamber
experiment~\cite{ref:camerini,ref:ljung} at the Argonne Zero
Gradient Synchrotron that was sensitive almost entirely to pion
momenta below 200 MeV/c. This experiment achieved a 90\% confidence 
level (CL) limit on
the branching ratio of $5.7 \times 10^{-5}$, assuming a pure vector
spectrum for the $\pi^+$.  Limits 
of  $3.1 \times 10^{-5}$ and $2.3 \times 10^{-5}$  
were extracted under the
assumptions of tensor and scalar interactions, respectively. 

Some features of the bubble chamber experiment are notable.
The experiment relied on the positive $\pi^+$ identification by
observation of the $\pi\to\mu\to e$ decay 
chain ($\pi^+\to\mu^+\nu$ followed by $\mu^+\to e^+\nu_e \bar\nu_\mu$). 
Although no timing information 
was available,  
kinematic information (specifically the measured range of the $\pi^+$ and the angle
between the incoming $K^+$ and outgoing $\pi^+$)  was used to 
reject background due to 
$K^+$ decay-in-flight. Events were discarded 
that showed evidence of a $\pi^+$-nucleus interaction in the
form of a drastic change in ionization along the $\pi^+$ track 
or a kink in the $\pi^+$ trajectory. Photon detection with a stated
inefficiency of 0.02 was used to veto $\pi^0$ decay products and 
provided additional
background suppression.

There followed a series of scintillation counter experiments by a Chicago-Berkeley
group that included a measurement in the range 
$142.7\ {\rm MeV}/c < P_{\pi^+} <  200.9\ {\rm MeV}/c$~\cite{ref:cable}.
  This yielded a 90\%
CL upper limit on the branching ratio of $9.4 \times 10^{-7}$ assuming
a vector spectrum.   
Corresponding limits were also determined assuming a 
tensor spectrum, $7.7 \times 10^{-7}$, a scalar spectrum, $1.1 \times
10^{-6}$ and other possible shapes.
In contrast to the bubble chamber experiment, 
the counter experiment made use of a 
delayed coincidence of 3.3 ns between 
the stopped $K^+$ and the outgoing track 
to suppress beam-related background including $K^+$ decay-in-flight. 
A hermetic 4-$\pi$ sr photon detector 
$\sim\!10$ radiation
lengths (r.l.) thick (4.3 r.l. along the incoming beam channel) 
 achieved a measured inefficiency for $\pi^0$ detection of 
$<2.2\times10^{-5}$ at 90\% CL for identified $K_{\pi2}$ decays~\cite{ref:pang}. 
As with the bubble chamber experiment, the $\pi\to\mu\to e$ chain
was used for positive $\pi^+$ identification and 
the measured range of the $\pi^+$ provided the  kinematic information
used in the analysis. A subsequent experiment at KEK that probed the \pnnone\ 
region improved the detection and identification of the 
$\pi\to\mu\to e$ decay by using 500 MHz waveform digitizers~\cite{Asano:1981nh}.

The next attempt at a measurement in the \pnntwo\  
momentum region emerged out of the first
phase of the E787 experiment at Brookhaven National 
Laboratory~\cite{ref:787first}.  
The E787 detector utilized and built upon concepts from the earlier
experiments. 
This experiment
obtained a 90\% CL upper limit of $1.7 \times 10^{-8}$, assuming a
$V-A$  spectrum modified by a form factor obtained from $K^+\to\pi^0 e^+ \nu$ 
data~\cite{Mescia:2007kn}.  We henceforth refer to this form as
the ``standard model'' interaction.  E787 also
obtained limits of $1.4 \times 10^{-8}$ and $2.2
\times 10^{-8}$, respectively, assuming pure tensor and scalar
interactions using  \pnntwo\  data exclusively~\cite{ref:jroy}.
 Adding  \pnnone\ 
data, E787 improved the limits to $1.0
\times 10^{-8}$ and $1.8 \times 10^{-8}$, respectively~\cite{ref:pnn1_2}.

The second generation of this experiment improved the SM limit 
in the \pnntwo\ region to $4.2
\times 10^{-9}$\cite{ref:787second} and subsequently to $2.2 \times
10^{-9}$\cite{ref:pnn2_PRD}.  Assuming tensor and scalar interactions
E787 ultimately obtained limits of $1.8 \times 10^{-9}$ and $2.7
\times 10^{-9}$, respectively~\cite{ref:bipul}.

\section{The E949 detector}  

\subsection{Detector description} \label{sec:e949det}

The E787 detector was upgraded in 1999-2000 to create the successor
experiment E949~\cite{ref:e949proposal}.
An extensive and detailed description of experiment E949  has been provided 
elsewhere~\cite{ref:pnn1PRD}. In this Section we provide 
a summary description of the detector and emphasize the features
essential to the \pnntwo\  region. 

\begin{figure}[h]
  \includegraphics[width=\linewidth]{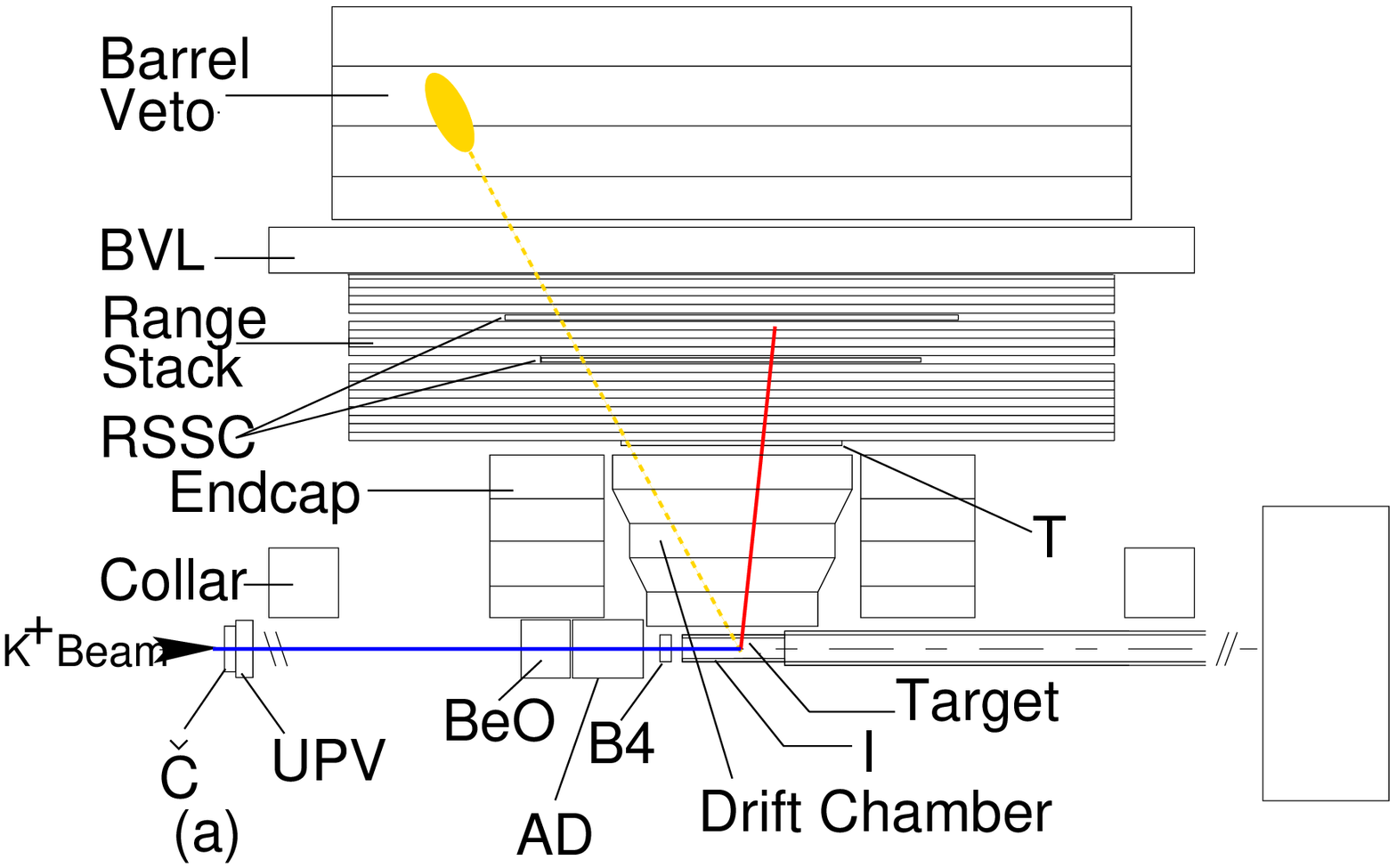} 
  \includegraphics[width=\linewidth,viewport=1 1 470 356]{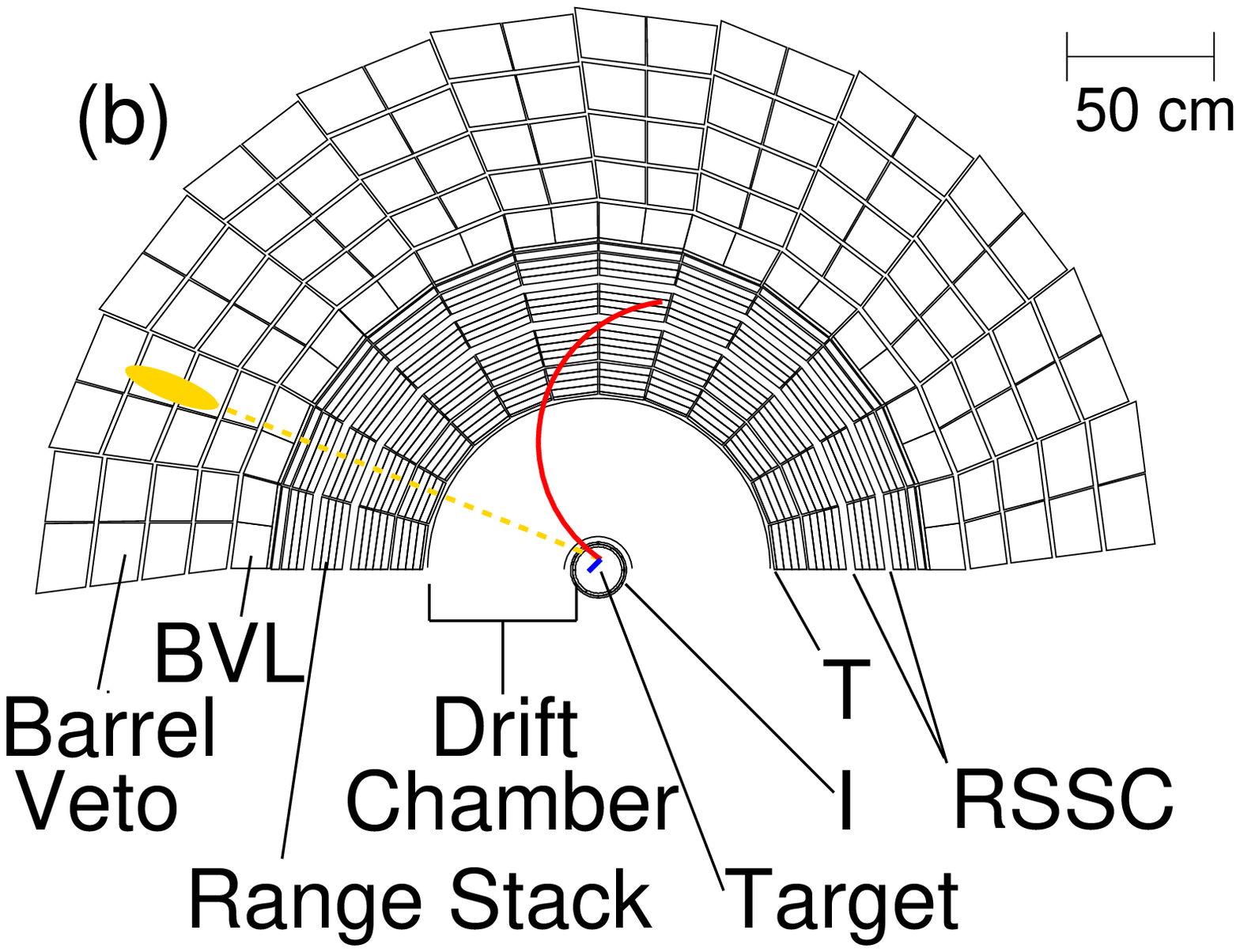} 
\caption{\label{fig:e949det}
Schematic side (a) and end (b) views of the upper half of the E949 detector.
An incoming $K^+$ is shown traversing the beam instrumentation, 
stopping in the target and decaying to $\pi^+\pi^0$. 
The outgoing charged pion and one photon from the 
$\pi^0\to\gamma\gamma$ decay are illustrated. 
Elements of the detector 
are described in the text.}
\end{figure}

E949 used an incident $710\ {\rm MeV}/c$ $K^+$ beam that was 
slowed and stopped in the scintillating fiber target as
shown schematically in Figure~\ref{fig:e949det}. Observation of
the decay \KPpnn\ requires detection of the incoming $K^+$ and
outgoing $\pi^+$ in the absence of any other coincident activity. 
The charged pion was identified kinematically by kinetic energy ($E_\pi$), 
momentum ($P_\pi$)  
and range ($R_\pi$) measurements and by observation of the $\pi\to\mu\to e$ 
decay sequence. Since the \KPpnn\ branching 
ratio was expected to be at the $10^{-10}$ level, the 
detector was designed to have powerful $\pi^+$ identification
to reject backgrounds from $K^+\to\mu^+\nu_\mu$ ($K_{\mu2}$), 
$K^+\to\mu^+\nu_\mu\gamma$ ($K_{\mu2\gamma}$) and $K^+\to\mu^+\pi^0\nu_\mu$ ($K_{\mu3}$), 
photon detection coverage over 4-$\pi$ solid angle to reject 
$K_{\pi2}$ and $K^+\to\pi^+\pi^0\gamma$ ($K_{\pi2\gamma}$), and 
efficient identification of a single incoming $K^+$ to 
suppress beam-related background.

The incoming charged-particle beam, containing approximately three 
$K^+$ for every $\pi^+$, traversed a \v{C}erenkov counter, 
two stations of beam wire proportional chambers (BWPCs), a passive BeO 
degrader, an active degrader (AD) and a beam hodoscope (B4) as 
shown in Figure~\ref{fig:e949det}.
The BWPCs, located between the UPV and BeO,
 are not explicitly shown in the figure.
 Typically, $1.6\times 10^6$ kaons per 
second entered the target during a 2.2 s spill. 
  \v{C}erenkov  photons emitted by an
incoming $K^+$ ($\pi^+$) passing through a lucite radiator 
were transmitted (internally reflected) into 14 ``kaon'' (``pion'') 
photomultiplier tubes to form  $C_K$ ($C_\pi$) 
coincidences.
The photomultiplier tube (PMT) signals were split and fed to a discriminator 
and a $\times 10$ amplifier. The discriminator output
was used as input to the time-to-digital converters (TDCs) 
and to the trigger (Section~\ref{sec:trigger}). The amplifier outputs 
were fed to 
500 MHz charge-coupled 
devices (CCDs)~\cite{ref:CCD}. 
The first (second) BWPC station was located downstream of the \v{C}erenkov 
counter at 168.5 (68.5) cm from the target entrance and each 
contained three planes with sense wires in the vertical and $\pm45^\circ$ ($\pm 60^\circ$) to 
the vertical direction. The wire spacing in the first (second) station
was 1.27 (0.80) mm. The BWPCs enabled detection of multiple beam particles.
The degraders were designed such that incident kaons stopped within the
fiducial volume  of the scintillating fiber target. 
The AD consisted of 40 layers of 2-mm thick plastic scintillator (13.9 cm diameter) 
interleaved with 39 disks of 
2.2-mm thick copper (13.6 cm diameter).  
The scintillator was azimuthally 
divided into 12 sectors that were coupled by wavelength-shifting (WLS) fibers 
to PMTs that were read out by analog-to-digital converters (ADCs), 
 TDCs and CCDs. These devices enabled measurement 
of activity in the AD coincident with the incoming beam and outgoing 
products of $K^+$ decays. The B4 hodoscope downstream of the AD
had two planes of 16 segmented plastic scintillator counters with 7.2-mm pitch 
oriented at $\pm33.5^\circ$ 
with respect to the horizontal direction. 
The cross-section of each counter was 
in a ``Z shape'' to minimize inactive area traversed by the beam and to improve
the spatial resolution~\cite{ref:Joss}. 
Each counter was connected to a PMT by three WLS fibers and
each PMT was read out by an ADC, a TDC and a CCD.  
The B4 enabled a measurement of 
the target entry position of the 
beam particle as well as identification of the
incident particle by energy loss. 

The target was composed of 413 scintillating 
fibers 3.1-m long with a 5-mm square
cross-section packed to form a 12-cm-diameter cylinder. A number of smaller 
(1-, 2- and 3.5-mm square) ``edge''  fibers filled the gaps at the outer edge
of the target. 
Each 5-mm fiber was connected to a PMT and the output PMT signal
was split and input into an ADC, a TDC, and low-gain($\times 1$) and 
high-gain($\times 3$) CCDs. 
The target fiber multiplicity and energy sum were also generated for
triggering purposes.
Multiple edge fibers were ganged onto 16 PMTs with similar readout.
Analysis of the 500 MHz sampling information provided by the target CCDs was 
essential
for isolating and suppressing backgrounds in the \pnntwo\ region. 
Two cylindrical layers of six plastic-scintillation counters 
surrounding the target defined the fiducial 
volume. 
The inner layer of counters (dubbed ``I counters'' or ``ICs'') 
were 6.4-mm thick with 
an inner radius of 6.0 cm and extended 24 cm from the upstream end of the
target. 
The 5-mm thick outer scintillation counters (VC) overlapped the downstream 
end of the ICs by 6 mm and extended 196 cm further downstream. The VC served
to veto particles that exited the target downstream of the IC.
Each IC and VC element was instrumented with a PMT and read out 
by an ADC, a TDC 
and a 500 MHz transient digitizer(TD)~\cite{ref:TD}. 

 The origin of the E949 coordinate system was the center of the cylindrical
volume defined by the ICs. This point also coincided with the 
center of the drift chamber. 
E949 employed a right-handed Cartesian
coordinate system with $+z$ in the incident beam direction, 
$+y$ vertically upward and the polar angle $\theta$ defined with respect 
to the $+z$ axis. The entire spectrometer was surrounded by 
a 1 T solenoidal 
magnetic field in the $+z$ direction.

The drift chamber, 
also called the ``ultra thin chamber'' (UTC)~\cite{ref:UTC}, 
was located just outside the IC, extended radially from 7.85 cm to 43.31 cm 
and served to measure the trajectory  
and momentum of the charged track from the target to the range stack
 as shown in Figure~\ref{fig:e949det}. 
Each of the three superlayers of the UTC contained four layers of axial anode
wires that provided \emph{xy} position information
and two cathode foil strips that provided \emph{z} position information. 
Beginning at an inner radius of 45 cm, the range stack consisted of 
19 layers of plastic scintillator counters and double-layer 
straw chambers (RSSC)~\cite{ref:mcpherson} embedded 
after the $10^{\rm th}$ and $14^{\rm th}$ layers of scintillator. 
The range stack enabled the measurement of the range and energy 
of the charged particle, 
the observation of the $\pi\to\mu\to e$ decay sequence and the measurement of 
photon activity. The 19 layers of plastic scintillator counters were
azimuthally segmented into 24 sectors as shown in Figure~\ref{fig:e949det}.
Layers 2--18 (19) were 1.9 (1.0)-cm thick and 182 cm long and were coupled 
on both ends to 
PMTs through  lucite light guides.
The trigger counters (T counters) in the innermost layer served to 
define the fiducial volume for $K^+$ decay products and 
were 6.4-mm thick and 52-cm long  counters coupled to PMTs 
on both ends by WLS fibers. The T counters were thinner than layers 2--19 to suppress 
rate  due to photon conversions. Signals from each range stack PMT were
passively split 1:2:2 for ADCs, discriminators and fan-in modules. The
discriminator outputs were sent to TDCs and used in the trigger. The fanned-in
analog sum of  four adjacent sectors 
(dubbed a range stack ``hextant'') was fed into 
a single TD and was provided to the trigger. The TDs digitized 
the charge in 2 ns intervals with an 8-bit resolution. The 500 MHz sampling 
permitted the observation of 
a $\pi^+\to\mu^+$ decay with a 5-ns separation between
the stopped pion and the emitted muon. 

\begin{figure}
\includegraphics[width=\linewidth]{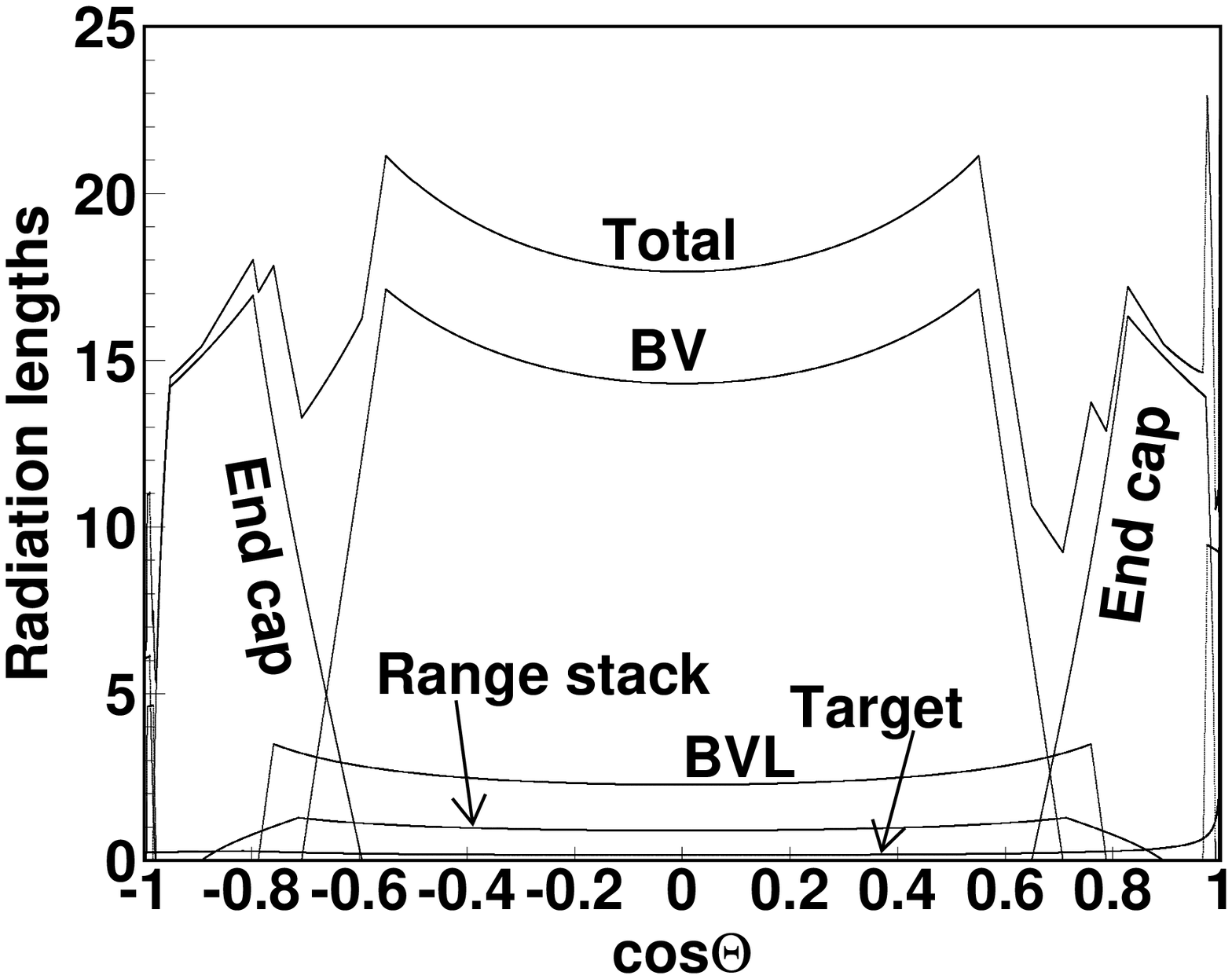}
\includegraphics[width=0.485\linewidth]{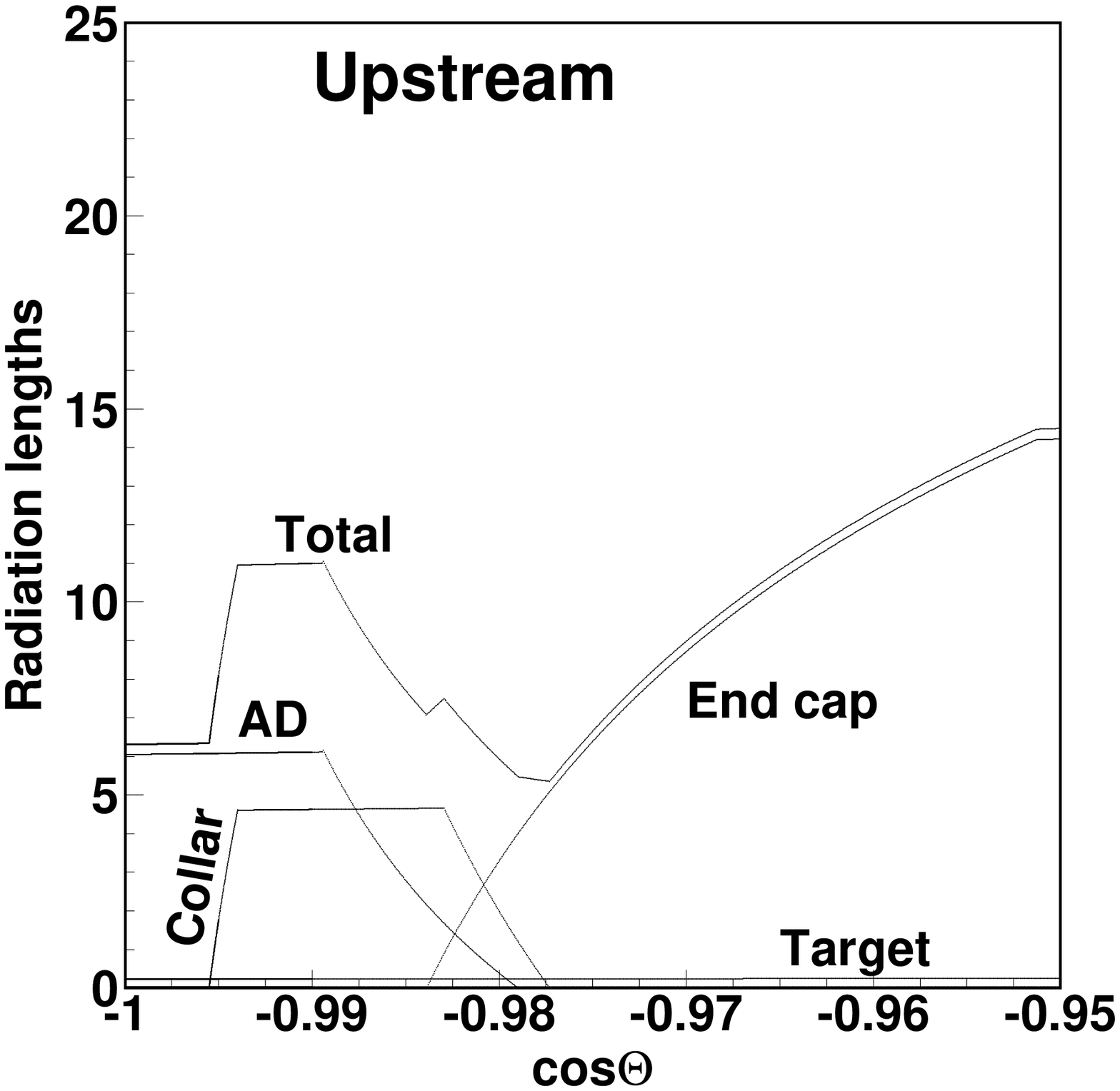}
\hskip -5mm
\includegraphics[width=0.485\linewidth]{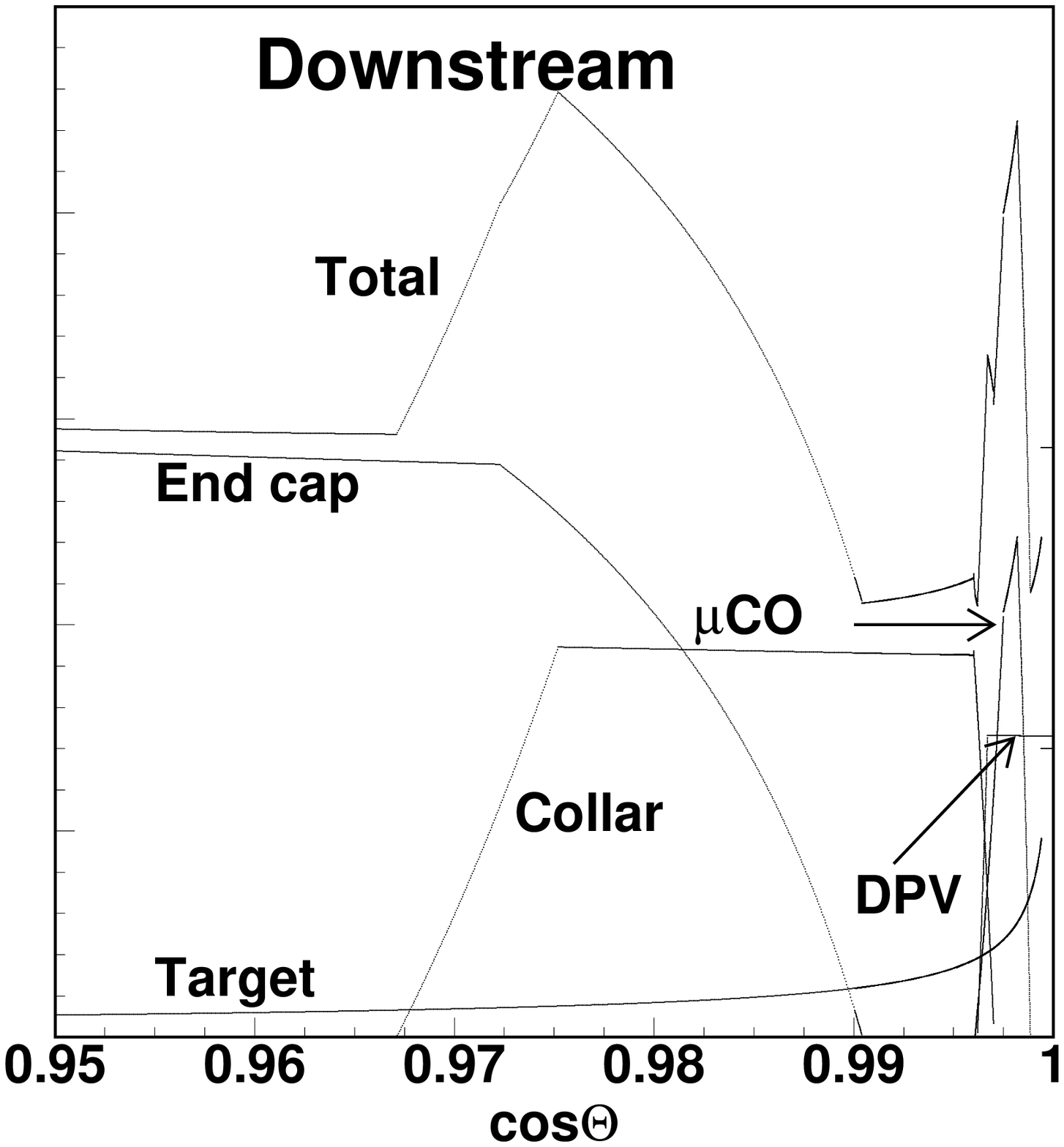}
\caption{\label{fig:pvthick} (Upper) Contribution of each photon veto detector 
in radiation lengths as a function of $\cos\theta$ for a photon emitted from the origin of the
detector coordinate system. 
(Lower) Contribution of each photon veto element in the upstream and downstream regions
within  $18^\circ$ of the beam.}
\end{figure}

 Identification of \KPpnn\ decays required detection of any activity
coincident with the charged track. Photons from $K_{\pi2}$ and radiative kaon
decays were detected in a hermetic photon veto system with 4-$\pi$ sr solid
angle coverage as shown in Figure~\ref{fig:e949det}. 
 Except for the end caps, 
all photon veto detectors were lead-scintillator 
sandwich-style electromagnetic 
calorimeters. 
Other detector elements, such as the range stack, target and AD, 
also served as  photon veto detectors. 
The barrel veto (BV)
and barrel veto liner (BVL) covered 2/3 of 4-$\pi$ sr 
outside the 
range stack with a radial thickness 14.3 and 2.29 r.l., respectively. 
The downstream and upstream end caps (ECs) consisted of
13.5 r.l. thick undoped CsI crystals and covered approximately 
1/3 of 4-$\pi$ sr~\cite{Chiang:1995ar,Komatsubara:1997rq}. The  3.1 r.l. thick 
upstream photon veto was mounted just downstream
of the \v{C}erenkov counter. The upstream and downstream
collar (CO) counters shown in Figure~\ref{fig:e949det} provided approximately 
4.5 and 9 r.l. at normal incidence, respectively. 
An additional collar counter ($\mu$CO) was installed downstream of the 
downstream CO between the inner face of the magnet end plate and the target~\cite{ref:pnn1PRD}. 
The downstream photon veto (DPV) provided 7.3 r.l. of coverage
downstream of the target, EC and collar. The AD was 6.1 r.l. thick and 
contributed important photon veto coverage in the poorly instrumented region near 
the beam axis. The thickness in radiation lengths of the photon veto system as
a function of the cosine of the polar angle is shown in 
Figure~\ref{fig:pvthick}. 
 
\subsection{Trigger}
\label{sec:trigger}  

 The trigger system for E949 was designed to select \KPpnn\ events from 
the large number of $K^+$ decays and scattered beam particles by requirements 
on the $\pi^+$ range, evidence of a $\pi^+\to\mu^+\nu_\mu$ decay in the range
stack, absence of other detector activity at the time of the $\pi^+$ and the 
presence of a  preceding $K^+$. The elements and architecture 
of the two-stage trigger system have been described previously~\cite{ref:pnn1PRD}; 
here we describe the features essential for  the \pnntwo\ region. 

 The logical OR of the following two signal triggers was used for the \pnntwo\ analysis,
\begin{eqnarray}\label{eqn:pnn1trig}
{\rm TRIG}\pnnone 
&\equiv& KB  \cdot (T\cdot2 \cdot IC) \cdot DC \cdot (6_{ct} + 7_{ct})\cdot \overline{19_{ct}} \nonumber \\
         & & \cdot \overline{zfrf} \cdot L0rr1 \cdot HEX \nonumber \\
         & &\cdot \overline{(BV+BVL+EC)}\cdot L1.1 \cdot L1.2
\end{eqnarray}

\begin{eqnarray}\label{eqn:pnn2trig}
{\rm TRIG}\pnntwo &\equiv& KB \cdot (T\cdot2\cdot IC) \cdot DC \cdot 3_{ct} \cdot 4_{ct} \cdot 5_{ct} \cdot 6_{ct} \nonumber \\
              && \cdot \overline{(13_{ct} + \cdots 18_{ct})} \cdot \overline{19_{ct}} \cdot L0rr2  \cdot HEX\nonumber \\
              && \cdot \overline{(BV+BVL+EC)} \cdot L1.1 \cdot L1.2 \ \ .
\end{eqnarray}                  
We collectively refer to the OR of the {\rm TRIG}\pnnone\ and {\rm TRIG}\pnntwo\ triggers as
{\oneortwo}.

 The $K^+$ beam condition $KB$ required a coincidence of at least five $C_K$ PMTs, 
the B4 hodoscope and the target with at least 20 MeV of deposited energy.   
The $KB$ signal served as the beam strobe for the trigger.
$T\cdot2\cdot IC$ required a coincidence  of the first two range stack 
layers in the same sector with
at least one IC to ensure that a charged track exited the target and entered the range stack.
The delayed coincidence ($DC$) required the IC time to be at least 1.5 ns later than 
the $C_K$ coincidence to select kaon decays at rest.
The ``$ct$'' designation refers 
to the range stack $T\cdot2$ sector and the next two adjacent sectors that would be 
traversed by a positively charged particle in the magnetic field.
 For the {\rm TRIG}\pnntwo\ trigger, 
the charged track requirements  
$3_{ct} \cdot 4_{ct} \cdot 5_{ct} \cdot 6_{ct} \cdot \overline{(13_{ct} + \cdots 18_{ct})} \cdot \overline{19_{ct}}$ 
ensured hits in range stack 
layers T through 6 to suppress contributions from 3-body $K^+$ decays 
and vetoed on hits in the outer layers to suppress long-range charged tracks beyond the \pnntwo\ kinematic region.
 The ${\rm TRIG}\pnnone$ trigger condition 
$\overline{zfrf}$  required the $z$ position of the charged track to be within
the fiducial region of all traversed range stack layers. 
The $L0rr1$ and $L0rr2$ were refined 
requirements of the charged track range taking into account the number of 
target fiber hits 
 and the track's $z$ position in range stack 
layers 3, 11, 12, 13 as well as the deepest layer of penetration
in order to reject long range tracks such as the $\mu^+$ from 
$K^+\to\mu^+\nu_\mu$  decay. 
The $BV$, $BVL$, $EC$ and $HEX$ requirements vetoed events with photons 
in the BV, BVL, EC and range stack, respectively. 
The $L1.1$ used 
the ratio of the height and area of the pulse(s) recorded by the TD
to select the two-pulse signature of the $\pi^+\to\mu^+$ decay in 
the range stack counter in which the charged track was determined to have stopped. 
The $L1.2$ used data digitized by the range stack ADCs to 
reject events with hits near the stopping counter that could falsely satisfy the $L1.1$ 
and to reject events with hits in both of the two adjacent hextants when the $T\cdot2$ and 
stopping counter were in the same sector.  
For the final 60.6\% of the data taking, an online pion \v{C}erenkov veto was 
included in the \pnntwo\ trigger to mitigate the effect of an increased
pion flux caused by reduced electrostatic separator voltage~\cite{ref:pnn1PRD}. 

\begin{figure}[h]
\includegraphics[width=\linewidth]{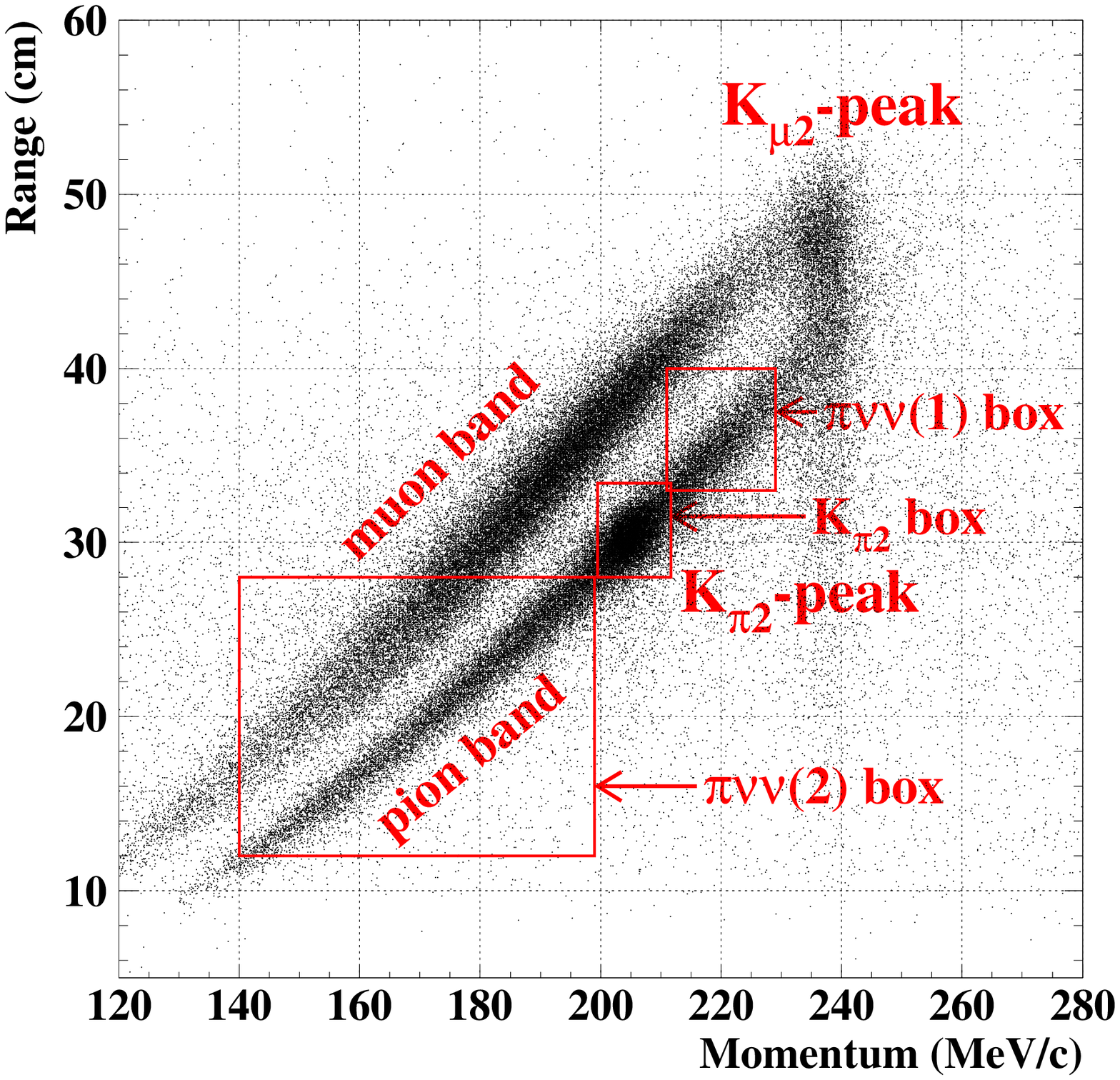}
\caption{\label{fig:triggers_rp} Range in plastic scintillator {\sl vs.} momentum for 
charged particles accepted by the \oneortwo\ 
trigger. The concentrations of events 
due to the two-body decays are labeled $K_{\pi2}$-peak and $K_{\mu2}$-peak.  
The decays $K^+\to\pi^0\mu^+\nu_\mu$ and $K^+\to\mu^+\nu_\mu\gamma$ contributed
to the muon band. The pion band resulted from $K^+\to\pi^+\pi^0\gamma$ decays, 
$K_{\pi2}$ decays in which the $\pi^+$ scattered in the target or range stack 
and beam $\pi^+$ that scatter in the target. 
The boxes at low and high momentum  represent the signal regions for this analysis 
and the previous \pnnone\ analysis~\cite{ref:pnn1PRD}, respectively. 
This distribution represents 0.13\% of the total kaon exposure.}
\end{figure}

A subset of the data selected by the \oneortwo\ trigger is shown in Figure~\ref{fig:triggers_rp}.

In addition to {\rm TRIG}\pnntwo\ and {\rm TRIG}\pnnone,  
 ``monitor'' triggers were
formed for calibration,  monitoring, and 
acceptance and background measurements~\cite{ref:pnn1PRD}. 
The monitor triggers selected events due to $K_{\mu2}$ and $K_{\pi2}$ decays as well as 
scattered beam pions ($\pi_{\rm scat}$).  An additional ``CEX'' 
monitor trigger requiring two $T\cdot2$ hits was  used to collect events 
resulting from the charge-exchange process $K^+n\to pK^0_S$ followed by 
$K^0_S\to\pi^+\pi^-$. Information derived from this CEX monitor data was used as input 
to simulation to determine the background from kaon charge-exchange reactions 
as described 
in Section~\ref{sec:CEX_bkgd}. In order to measure the efficiency of 
the $T\cdot2\cdot IC$ condition (Section~\ref{sec:acceptance}), 
we also defined a KB monitor trigger that 
required the $KB$ condition described previously. 
All monitor triggers were prescaled to reduce their contribution to
the deadtime.

\section{Data Analysis}
\label{sec:Data_Analysis}  

The total exposure for this analysis was \KBLIVE\ stopped kaons corresponding 
to $1.43\times10^8$ \oneortwo\ triggers. The total exposure was
slightly less than the $1.77\times10^{12}$ stopped kaons used for the  \pnnone\ 
analysis~\cite{ref:pnn1PRD} because some data were discarded due to more 
stringent requirements on the reliability of the
BWPCs, the \v{C}erenkov counter and the target CCDs.

\subsection{Overview}
\label{sec:Overview}  

 Identification of the \KPpnn\ decay involved positive observation of 
the $K^+$ and daughter $\pi^+$ in the absence of coincident  detector
activity. 
The \pnnone\ and \pnntwo\ regions in E949 extended 
from 211 to 229 ${\rm MeV}/c$~\cite{ref:pnn1PRD} 
and 140 to 199 ${\rm MeV}/c$ in $\pi^+$ momentum 
below the $K_{\pi2}$ peak, respectively (Figure~\ref{fig:triggers_rp}).

The \pnntwo\ region potentially has a  larger 
acceptance than \pnnone\ because
the phase space is larger and the loss of $\pi^+$ due to nuclear 
interactions in the detector 
is smaller at lower pion energies. 
These factors partially mitigated the 
loss of acceptance due to additional requirements needed to suppress background 
in the \pnntwo\ region. 
Compared to the previous \pnntwo\ analyses~\cite{ref:pnn2_PLB,ref:pnn2_PRD},
the acceptance was increased by enlarging the size of the signal
region.

In a further enhancement to the previous \pnntwo\ analyses~\cite{ref:pnn2_PLB,ref:pnn2_PRD}, 
the signal region was sub-divided into regions with differing signal-to-background 
ratios. The signal-to-background of each region was taken into account 
in the evaluation of ${\cal B}(\KPpnn)$ 
using a likelihood method (Section~\ref{sec:like}).

\subsubsection{Kaon-decay background}

In the \pnnone\ 
region, the background was dominated by $K_{\pi2}$, 
$K_{\mu2}$, $K_{\mu2\gamma}$ 
and $K_{\mu3}$ decays and was sufficiently suppressed by 
positive identification 
of the $\pi^+$ based on kinematic properties,  observation of the 
$\pi\to\mu\to e$ sequence 
and by the hermetic photon veto capability~\cite{ref:pnn1PRD}. 
Previous studies~\cite{ref:pnn2_PLB,ref:pnn2_PRD} 
in the \pnntwo\ region identified the main background as due to $K_{\pi2}$ 
decays in which the charged pion 
scattered in the target, lost energy and fell into the signal region.
The scatter reduced the 
directional correlation between the charged and neutral pions.
Thus the photons from $\pi^0$ decay were directed 
away from the high efficiency barrel region of the photon veto.
This background was suppressed, in part,  by recognition of the scattering process
in the target. A background contribution due
to scattering of the charged pion in the range stack was suppressed 
by the track pattern and energy deposit in the range stack.
The photon veto served to suppress these ``$K_{\pi2}$-scatter'' backgrounds 
as well as background due to the radiative  decay $K_{\pi2\gamma}$.
Background due to  $K^+\to\pi^+\pi^-e^+\nu_e$ ($K_{e4}$) was suppressed by identification of 
additional particles in the target. Kaon decays with a muon in the
final state ($K_{\mu2}$, $K_{\mu2\gamma}$ and $K_{\mu3}$)
were suppressed by kinematics and the recognition of the $\pi\to\mu\to e$ 
signature as well as the photon veto for the latter two decays.

\subsubsection{Beam-related background}

The beam-related backgrounds were categorized 
as being due to CEX, one beam particle (single-beam background), 
and two beam particles (double-beam background).
The CEX background occurred due the production of a $K^0$ 
in the target from the charge-exchange process $K^+n\to p K^0$. 
If the $K^0$ turned into a $K^0_L$ that subsequently underwent
semileptonic decay ($K_L^0\to\pi^+\ell^-\bar\nu$ with $\ell^-=e^-$ or $\mu^-$), the
$\pi^+$ could fall in the \pnntwo\ kinematic region. CEX background
was rejected by observing the gap between the kaon and pion fibers
due to propagation of the non-ionizing $K^0_L$, by the inconsisteny
between the energy deposited by the $K^+$ and the reconstructed
$z$ of the outgoing pion and by identification of the 
accompanying negative lepton. In addition, requirements on the delayed
coincidence between the $K^+$ and $\pi^+$ suppressed CEX background due
to the short $K^0_L$ flight time.

Single-beam background was  due to a $K^+$ entering the
target and decaying in flight to produce a $\pi^+$ in 
the \pnntwo\ region. Incoming beam $\pi^+$ misidentified as $K^+$ 
and scattering in the target also contributed to the 
single-beam background. Positive identification of the incoming particle 
as a kaon as well as requirements on the delayed coincidence between
the incoming and outgoing tracks suppressed the single-beam background.

The two processes (kaon decay-in-flight and pion scattering) 
that contributed to single-beam background formed the double-beam background
when preceded by an additional incoming kaon whose decay products were 
undetected. Double-beam background was suppressed by requiring an absence of
activity in the beam detectors 
in coincidence with the $\pi^+$ detected
in the range stack.

\subsubsection{Analysis method and strategy} \label{sec:analmeth}

We used  analysis procedures and strategies similar to that of 
the E949 analysis of the \pnnone\  region~\cite{ref:pnn1PRD} 
with some modification 
 that took into account the difficulty of isolating 
some background samples in the data in the \pnntwo\ region. 
As with the previous analysis, we adopted a ``blind'' analysis method in that 
we did not examine the pre-defined signal region until all background
and acceptance analysis was completed. Since we also attempted to obtain all 
background estimates directly from the data, we inverted at least one
selection criteria (``cut'') when we used the \oneortwo\ data to avoid
examining the signal region.  Every 
third \oneortwo\ trigger formed the ``1/3'' sample that was used to 
determine the selection criteria. 
We then obtained unbiased background estimates by applying the finalized 
selection criteria to the remaining ``2/3'' sample of \oneortwo\ triggers.

\begin{figure}[h]
\includegraphics[width=\linewidth]{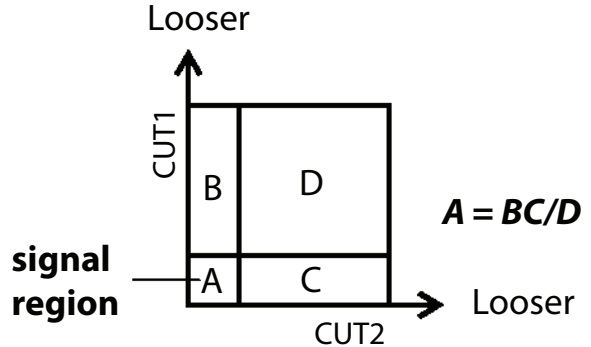}
\caption{\label{fig:bifurc}Schematic of the bifurcation method.
The background level in region $A$ can be estimated from the 
number of events observed in  regions $B$, $C$ and $D$ assuming CUT1
and CUT2 are uncorrelated. See text for details.}
\end{figure}

The preferred method of background estimation employed the 
bifurcation method illustrated in Figure~\ref{fig:bifurc}.
The parameter space of two sets of uncorrelated cuts ``CUT1'' and ``CUT2'' can be
divided into the four regions shown in the figure by the application
of each cut or the inverted cut. The number of 
events in the signal region ``$A$'' can be
determined by application of both CUT1 and CUT2.
If the background rejection of CUT1 was independent of CUT2, 
then the number of events in $A$ can be estimated as the number
of events in region $B$ times the ratio of the number of events in 
regions $C$ and $D$ or $A=BC/D$. 
In practice, we employed two branches for the bifurcation
analysis. The ``normalization branch'' analysis was performed to 
obtain the number of events, $N_{\rm norm}$,  
in region $B$. A ``rejection branch'' analysis
was used to obtain $D/C$. We defined the rejection as $R\equiv (C+D)/C$ and
obtained the background estimate as 
\begin{equation}\label{eqn:bif}
b = f\times N_{\rm norm}/(R-1)
\end{equation}
where $f = 3(3/2)$ for the 1/3(2/3) sample. 
For all background estimates in this analysis, the normalization 
branch was taken from the \oneortwo\ data. We used the \oneortwo\ data 
to obtain the rejection branch for all backgrounds except 
for the CEX, $K_{e4}$ and $K_{\pi2\gamma}$ backgrounds that could not
be cleanly isolated in data. For these backgrounds, EGS4-based simulations~\cite{ref:EGS} 
 were employed. 
When no events ($N_{\rm norm}=0$) were available in the 
normalization branch, we 
assigned $N_{\rm norm}=1$.

We checked the validity of the background estimates by loosening cuts
and comparing the predicted 
 number of events just outside the signal region with 
observations (Section~\ref{sec:otb}). 
In addition we examined events failing only a single major selection criteria
to search for unforeseen background sources and coding 
mistakes (Section~\ref{sec:singlecut}).
\subsection{Data selection} \label{sec:requirements}
\subsubsection{Event reconstruction}\label{sec:eventrecon}

Event reconstruction was performed in a number of steps consisting of track-fitting in 
various detector systems such as the beam-line detectors, the range stack, the UTC
and the target.
Multiple iterations of the track-fitting were performed in many of the detector 
systems using
progressively better information from track-fitting from other detector systems as
constraints.
Events were reconstructed as described in \cite{ref:pnn1PRD} except as noted
below. 

The following discussion focuses on the target track-fitting to clearly define the 
target-fiber classification scheme for use in the description of the target CCD fitter and
 the cuts that used target fiber information. 
In contrast to the analysis of the \pnnone\ region~\cite{ref:pnn1PRD}, the 
fit to the UTC track did not include information from the target fibers. 
Performing the target fit separately improved the ability to detect 
a pion scatter in or near the target.

\begin{figure}
        \includegraphics[width=\linewidth]{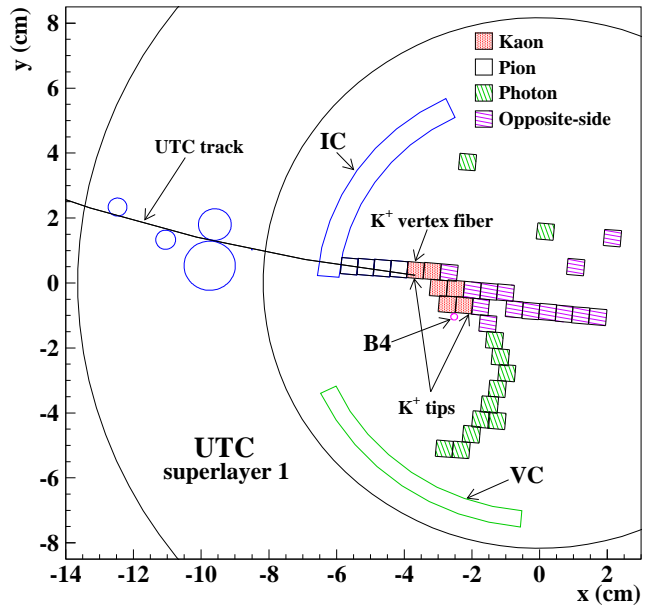}
        \caption{\label{fig:target1}Example target \emph{xy} display showing
        the assignment of pion, kaon, photon and opposite-side pion fibers
        as described in the text. The arc that terminates in the reconstructed
        kaon vertex fiber represents the extrapolated UTC track. The
        circles on the track represent the UTC hits with the radius giving
        the drift distance. Only the innermost superlayer of the UTC is shown.
        The IC and VC elements with hits are indicated.
        The position of the incoming $K^+$ as reconstructed by the B4 hodoscope
        information is also indicated in the figure as are 
        the kaon fiber ``tips'' described in Section~\ref{sec:target_cuts}.
        This event was selected in the $K_{e4}$ 
        normalization branch (Section~\ref{sec:Ke4_bkgd}).
        The measured time, energy and apparent curvature 
        of the contiguous fibers classified as ``photon'' fibers 
        are consistent with a positron
        and the time and energy of the opposite-side pion fibers are
        consistent with a negative pion.}
\end{figure}

\begin{figure}
  \includegraphics[width=\linewidth,height=0.30\textheight]{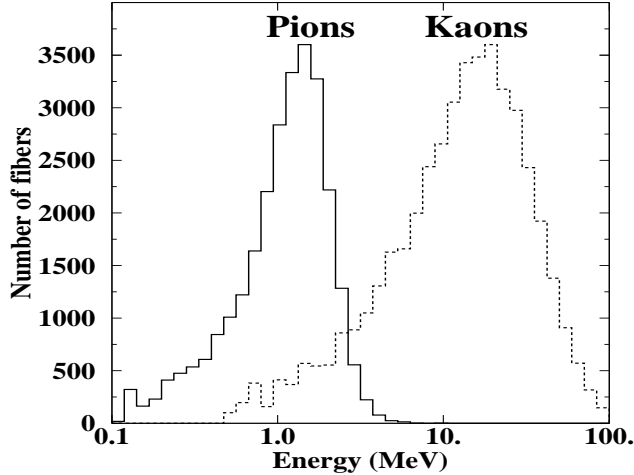}
  \caption{\label{fig:epi_ek}The energy per pion fiber or kaon fiber 
    in events in the $K_{\pi2}$-peak region, defined in Section~\ref{sec:kinematic_cuts},  
    in \oneortwo\ triggers.
    The height of the pion energy distribution was normalized to
    that of the kaon energy distribution for the purpose of display.
    The average number of pion (kaon) fibers per event in the selected events
    is 10.2 (5.3).
    Note the logarithmic abscissa.}
\end{figure}

After the range stack and UTC track fitting were performed,
target fibers were clustered into $K^+$ and $\pi^+$ paths
based on geometry, energy and timing information 
as shown in Figure~\ref{fig:target1}.
The pion fibers had to lie along
a strip (typically 1 cm in width) along the UTC track extrapolated
into the target, have an energy between 0.1 and 10.0 MeV and be
in coincidence with the reconstructed time of the $\pi^+$ in the 
range stack ($t_{rs}$). For the first iteration, the kaon fibers had to 
have greater than 4 MeV of energy and be coincident with the beam strobe time. 
In subsequent iterations, fibers of lower energy which were contiguous
with the putative kaon track could be classified as kaon fibers.
Any fiber that did not fall into the kaon or pion  fiber categories
was classified as a photon  fiber if it had more than 0.1 MeV of energy.
The $K^+$ decay vertex fiber was identified as the kaon  fiber closest to the
extrapolated UTC track and farthest from the \emph{xy} position
of the B4 hit.
Hit fibers that were located on the opposite side of the decay vertex with
respect to the outgoing track were classified as ``opposite-side pion'' fibers 
and were possibly due to $K^+$ decays with multiple charged particles or a photon
conversion.
The energy-weighted average times of the $K^+$ and $\pi^+$ hits were 
 ${t_K}$ and ${t_{\pi}}$, respectively.
The sums of the $K^+$ and $\pi^+$ energies were 
${{E_K}}$ and ${{E_{\pi}}}$, respectively. 
The pion and kaon fiber energy distributions 
are shown in Figure~\ref{fig:epi_ek}.

Identified pion fibers were subjected to a least-squares fit to 
the hypothesis of a positively charged pion track~\cite{ref:bipul}. 
The ``target-track fitter'' tested the consistency of the 
energy-loss-corrected UTC track with the energy in the fibers 
or the distance to the track if the fiber was not on the track.

To aid in the identification of pion scattering in the target, 
the activity in each of the target fibers
as recorded by the low-gain and high-gain CCDs (Section~\ref{sec:e949det}) 
was fitted using a single-pulse and a double-pulse hypothesis.
For each CCD on each fiber,
a standardized $K^+$ pulse used for the target CCD fitter was created using
kaon fiber data from $K_{\mu2}$ monitor trigger data.
For each fiber having an energy greater than a fiber-dependent threshold,
typically 2 (0.5) MeV for low (high) gain, 
the fitting procedure was performed on the low-gain and the
high-gain CCD information independently.
The first step of the procedure was a least-squares fit
 to a single-pulse hypothesis 
  for each fiber channel
passing the above criteria.
The single-pulse fit used two parameters, the pulse
amplitude and the time.
If the probability of $\chi^2$ (${\cal P}(\chi^2)$)  
of the single-pulse fit was less than 25\%,
a double-pulse fit was performed.
The double-pulse fit used four parameters, the amplitudes and
times for the first and second pulses.

\subsubsection{Requirements on $\pi^+$ in  target}
\label{sec:target_cuts}  

 Numerous requirements were placed on the activity in the
target to suppress background and ensure reliable determination 
of the kinematic properties of the charged pion.
These requirements were based on the results of the target
CCD fitter, the reconstructed energy and time of the pion and 
kaon fibers, the pattern of kaon and pion fibers relative to 
information from the rest of the detector and the results
of the target-track fitter.

\subsubsubsection{Target-pulse data analysis} 

\begin{figure}[h]
  \includegraphics[width=\linewidth]{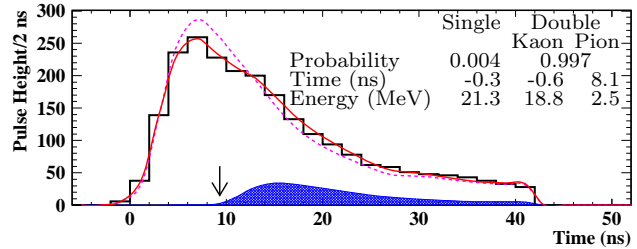}
  \vskip -3.7cm
  { \footnotesize \hfill
    \begin{tabular}{lrrr}
                &Single & \multicolumn{2}{c}{Double}\\ [-0.8mm]
                &       & Kaon        &  Pion       \\ [-0.8mm]
    Probability &0.004  & \multicolumn{2}{c}{0.997} \\ [-0.8mm]
    Time (ns)   & -0.3  & -0.6        & 8.1         \\ [-0.8mm]
    Energy (MeV)& 21.3  & 18.8        & 2.5         \\ [-0.8mm]
    \end{tabular}
    \hspace{0.1cm}
    }
  \vskip  1.8cm 
  \caption{\label{fig:ccdfit}
       CCD pulse fit example. The histogram represents the
    pulse height distribution for the high gain CCD data. The 
    histogram was terminated at 42 ns due to a software cutoff.
    The dashed (solid) line represents the fitted total pulse shape for 
    the single-(double-)pulse hypothesis. The filled area represents the
    fitted second pulse for the double-pulse
    hypothesis. The arrow indicates $t_{rs}$, the expected time of the second
    pulse based on the reconstructed $\pi^+$ in the range stack.
    This event was rejected because the 2.5 MeV of the fitted second pulse was 
    coincident with $t_{rs}$. 
  }
\end{figure}

 Detection of pion scattering in the target in the identified kaon 
fibers required reliable
results from the target CCD fitter. 
An algorithm, 
based on the energy in the kaon fiber as
measured by the ADC and by the time difference $t_\pi - t_K$,
determined if the information from the high-gain CCD, the low-gain CCD or 
a combination of the two should be used for each fiber with an acceptable
double-pulse fit. 
If ${\cal P}(\chi^2)$s of
the fits for both the single- and double-pulse hypotheses were less than
 $5\times 10^{-5}$ 
 in any of the kaon fibers, then 
the event was rejected. In addition, the fitted 
time of the first pulse ($t_1$) was required to be consistent with the average
time of the kaon fibers. If any kaon fibers failed the requirement 
$-6 < t_1 - t_K < 7 $ ns, 
then the event was rejected. This requirement was made on the fitted time $t_1$ of
the single-pulse hypotheses if the probability of $\chi^2$ was greater than 25\% and
on the fitted time $t_1$ from the double-pulse hypothesis otherwise. 
The requirement on $t_1$ rejected events in which the CCD 
fitter attempted to fit a fluctuation
in the tail of 
the data pulse or when there was a large
second pulse 
and the fitter mistakenly identified it as the 
first pulse. For events passing these criteria,
the second-pulse activity in a kaon  fiber as found by the target
CCD fitting was required to be below 1.25 MeV
when the fitted second-pulse time $t_2$ satisfied the
coincidence condition $-7.5 \leq t_2 - t_\pi \leq 10$ ns. 
An example of the fit for a high gain CCD target element is shown 
in Figure~\ref{fig:ccdfit}. 
In the following we refer to these requirements on the CCD pulse fitting
as the ``CCDPUL'' cut. 

\subsubsubsection{Kaon fiber timing}

The target  kaon fiber hits
were required to be consistent with a kaon approaching the $K^+$ decay vertex.
This consistency was enforced by requiring ${\cal P}(\chi^2)$ to be
greater than 5\% for 
fits to the kaon fiber hit times {\it vs.} {\emph {xy}} distance to the decay vertex 
and {\it vs.} range. 
This requirement removed events in which the kaon decay vertex was 
incorrectly assigned. 

\subsubsubsection{Pion fiber energy}

Pion fibers were required to have
energies less than 3.0 MeV.
This suppressed $\pi^+$ target-scatters
since the expected mean energy deposited in a  pion fiber was 
approximately 1.2 MeV.  This cut had an acceptance factor of
89.6\% (Section~\ref{sec:acckp2}) due to the Landau distribution that
describes the ionization energy deposit. 

The measured range and energy of the pion and the pion momentum were required to
be consistent with that expected for a $\pi^+$ using a cut on a likelihood 
function. The likelihood function was calibrated using $\pi_{\rm scat}$ monitor
trigger events.
In addition,  the total energy of the pion  target fibers 
was required to 
be in the range of 1 to 28 MeV and 
the total energy within $\pm4.0$~ns of $t_{rs}$ in the  target 
edge fibers was required to 
be less than 5.0 MeV.

\subsubsubsection{Pattern of kaon and pion fibers}

Events with a minimum distance between the centers of the closest pair of kaon
and pion fibers greater than 0.6 cm, more than one fiber width, were
rejected. This cut suppressed the CEX background. 
A more stringent version of this cut that also required that no 
photon fibers filled the gap between the kaon and pion fibers
was developed to define the normalization branches for the CEX 
 (Section~\ref{sec:CEX_bkgd}) and double-beam (Section~\ref{sec:doublebeam}) 
background measurements.


Two conditions were used to enforce consistency among the positions of
the  kaon decay vertex, the kaon and pion clusters, and the beam particle
in the B4 hodoscope.
The first condition required that the distance in the \emph{xy}-plane
between the hit position in the B4 hodoscope and the nearest tip of the 
kaon fiber cluster be less than 1.8 cm. 
The kaon  cluster tips were defined to be 
the two kaon fibers farthest apart from each other (Figure~\ref{fig:target1}). 
The second condition required that 
the distance in the \emph{xy}-plane
between the kaon decay vertex 
and the nearest kaon cluster tip was
less than 0.7 cm. 
This requirement suppressed $K_{\pi2}$ target-scatter background 
when the scattered $\pi^+$ did not emerge from the 
fiber containing the $K^+$ decay. 

The total energy of opposite-side pion fibers  
within $\pm4.0$ ns of $t_\pi$ was required to be less than 1.0 MeV to 
suppress background due to $K_{e4}$ decays as well as $K_{\pi2}$ scatters.
Hereafter, this cut is referred to as ``OPSVETO''.

\subsubsubsection{Target-track fitter}

The track determined by the target-track fitter
was required to be consistent with the information
in the target fibers and the fitted UTC track 
in order to suppress backgrounds due to pion scattering, CEX, $K_{e4}$ or a second
beam particle in the target.
For three contributions 
$\chi^2_5$, $\chi^2_6$ and $\chi^2_7$ to  the $\chi^2$ 
for the target-track least-squares fit, 
${\cal P}(\chi^2_5+\chi^2_6+\chi^2_7)$ was required to be greater than 
1\%. These were defined as follows:
\begin{itemize}
\item[$\chi^2_5$] was assigned a contribution for each pion fiber
traversed by the track based on the comparison 
of the observed energy with the expected
energy from  the calculated range of the track and the track momentum.
\item[$\chi^2_6$] was assigned a contribution based on the minimum distance 
between the track and the nearest point of each
fiber that was traversed by the track, but had no observed energy. 
This assignment acted to force the fitted track to go between fibers and
thus provided precise position information on the track.
\item[$\chi^2_7$] was assigned a contribution for pion fibers that were not
traversed by the fitted track based on the distance to the nearest corner.
\end{itemize}
Events were rejected if any single pion fiber contributed 
more than 35 units to $\chi^2_5$ which might indicate a pion scatter in that
fiber.
The fitted target track was also required 
to 
intersect the kaon  vertex fiber. 
The angle between the reconstructed target track  and the 
UTC  track was required to be less than 0.01 radian 
at the radius of the IC 
when the range of the $\pi^+$ in the target was less than 2.0 cm. 
In addition, the position of the reconstructed $\pi^+$ trajectories from 
the target and UTC fits were required to be well-matched at the target edge.
Events with a kink in the target $\pi^+$ track were suppressed by requiring that
the difference in the distances in the {\em xy}-plane 
of the farthest and nearest pion fibers to the center of the
fitted helix of the UTC track was less than 0.35 times the pion range in the target.

\subsubsection{Pion track requirements}
\label{sec:track_cuts}  

Good pion track reconstruction was required based on 
the $\chi^2$ of the UTC track fit.
The cut  on the $\chi^2$ was dependent on the number of
anode and cathode hits assigned to the fitted track 
as well as  on the number of unused
anode and cathode hits. The criteria were determined using 
both  $K_{\pi2}$  and $\pi_{\rm scat}$ monitor trigger data 
such that the $\pi^+$ momentum resolution of $2.3\ {\rm MeV}/c$  for the
$K_{\pi2}$ peak~\cite{ref:pnn1PRD} was 
maintained while retaining high efficiency.

Range-stack quality cuts  were placed on the probability of $\chi^2$ of
the range-stack track fit and the agreement of the
$z$ position of the extrapolated UTC track 
with the range-stack timing information and, when applicable, the 
RSSC information. The RSSC was not available for 
charged particles  that stopped in range stack layers 6 through 10.

\begin{figure}[h]
        \includegraphics[width=\linewidth,height=0.30\textheight]{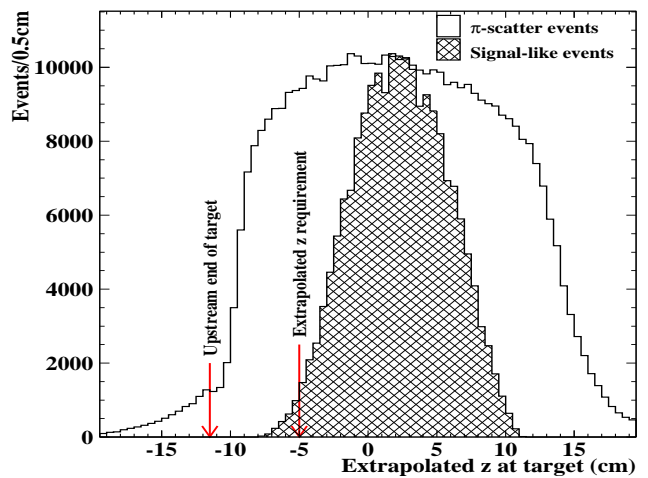}
        \caption{\label{fig:tgz}
        Extrapolated target $z$ distribution
        of the charged track. 
        The ``$\pi$-scatter'' and ``signal-like'' events are taken from
        $\pi_{\rm scat}$ and $K_{\mu2}$ monitor trigger data, respectively.
        The  required minimum  on the extrapolated $z$ position 
        and the upstream end of 
        the target are indicated  in the figure.     } 
\end{figure}
Kaons that came to rest in the target were required to 
have  energy loss in the B4 hodoscope and the target consistent
with the measured $K^+$ stopping position based on the $z$ position
of the extrapolated UTC track.  A likelihood function 
based on these
three quantities, was calibrated using $K_{\mu2}$ monitor trigger data.
The requirement on this ``Beam Likelihood'' function  suppressed
$K_{\pi2}$ target-scatter and CEX backgrounds as well as background
due to an incoming beam pion that scattered in the target.
In addition, we required the $z$ of the extrapolated UTC track to be
greater than $-5.0$ cm  as 
shown in Figure~\ref{fig:tgz}. 
Since other cuts more effectively suppressed background from 
the downstream portion of the target ($10< z < 20\ {\rm cm}$), 
no explicit additional  cut on $z$ was necessary.

\subsubsection{Decay pion kinematic requirements}
\label{sec:kinematic_cuts}  

The total range (kinetic energy) of the $\pi^+$ track was
calculated as the sum of the measured range (energy) in 
the target, IC and range stack. The total momentum was
obtained from the curvature of the fitted track in the UTC 
corrected for 
 energy loss in the target and IC.
Tiny corrections were applied to $R_\pi$, $E_\pi$ and $P_\pi$ 
to take into account the inactive material in the UTC~\cite{ref:pnn1PRD}.

The upper limit of the signal region in range, energy and momentum
was increased with respect to the previous \pnntwo\ analyses~\cite{ref:pnn2_PLB,ref:pnn2_PRD}
to be approximately 2.5 standard deviations from the $K_{\pi2}$ peak 
similar to the approach used for the E949 \pnnone\ analysis~\cite{ref:pnn1PRD}.
The lower limits were not changed with respect to the previous \pnntwo\ analyses.
The standard signal region was 
$ 140 < P_\pi < 199\ {\rm MeV}/c$, 
$ 60  < E_\pi < 100.5\ {\rm MeV}$ and
$ 12  < R_\pi < 28\ {\rm cm}$.

To increase the statistical power of any observed signal events (Section~\ref{sec:like}), 
a tighter kinematic  region was defined as a subset of the standard
region to further suppress $K_{\pi2}$ and $K_{e4}$ backgrounds.
As shown in Figure~\ref{fig:ke4_norm},  $K_{\pi2}$ and $K_{e4}$ events were not
uniformly distributed in the signal region.
The $K_{\pi2}$ target-scatter events were more uniformly distributed in 
the signal region except near the $K_{\pi2}$ peak. 
The imposition of the \oneortwo\ trigger on the $K_{e4}$ momentum distribution 
shown in Figure~\ref{fig:seven} caused 
the $K_{e4}$ background to peak around 160 MeV/c as 
described in Section~\ref{sec:Ke4_bkgd}.
The accepted \KPpnn\ spectrum was monotonically increasing with momentum in the
signal region (Figure~\ref{fig:stx}).
Based on these observations, the 
kinematic region that maximized the ratio of the signal acceptance 
to  the total 
$K_{\pi2}$ and $K_{e4}$ background was 
$165 <  P_\pi <  197\ {\rm MeV}/c$,
$72 <  E_\pi <  100\ {\rm MeV}$ 
and $17 < R_\pi < 28\ {\rm cm}$.

We also defined the ``$K_{\pi2}$-peak region'' by the requirements
$199 < P_\pi < 215\ {\rm MeV}/c$, 
$100.5 < E_\pi < 115\ {\rm MeV}$ 
and $28 < R_\pi < 35\ {\rm cm}$ (Figure~\ref{fig:triggers_rp}). 
Events in the $K_{\pi2}$-peak region were employed to set selection 
criteria, estimate background and determine the signal 
acceptance.
\subsubsection{Muon identification}
\label{sec:muon_rej}

Muon backgrounds were rejected largely based upon
the positive identification of the $\pi^+$ in the
range stack by the observation of the $\pi\to\mu\to e$ 
decay chain and by the range-momentum relationship.

The $\pi^+$ identification algorithms of {\pnnone}~\cite{ref:pnn1PRD} 
were adopted for this analysis, 
and  only a brief description  is provided here.
The analysis of
the waveform provided by the transient digitizers (TDs) was
used to identify the $\pi^+\to\mu^+\nu_\mu$ decay in the 
range stack element that contained the stopping pion. A neural network
 was trained using kinematically 
identified $\pi^+$ and $\mu^+$ that stopped in the range stack~\cite{ref:pnn1PRD}.
The $\mu^+\to e^+\nu_e\bar\nu_\mu$ decay was identified by 
TDC information in the range stack counters near the stopping counter.

Since the previous \pnntwo\ analyses~\cite{ref:pnn2_PLB,ref:pnn2_PRD} had shown
that muon backgrounds were small, less restrictive requirements 
on the $\pi\to\mu\to e$ decay  than  those in the \pnnone\ analysis were 
used for the standard \pnntwo\ requirements.
This provided a 10\% increase
in the signal acceptance. A muon 
rejection of $133.0\pm10.7$ (Section~\ref{sec:Muon_bkgd}) 
was obtained with a looser 
requirement on the neural network output 
and no identification of the $\mu\to e$ decay.
The \pnnone\ requirements were used to define a tighter
cut that was used to subdivide the signal region as described
in Section~\ref{sec:like}. The tighter cuts had a
muon rejection of $409.1\pm60.9$. In the following we 
refer to the $\pi\to\mu\to e$ requirements as the ``TD'' cuts.

The ability to separate pions from muons using the range and momentum
measurements can be seen in Figure~\ref{fig:triggers_rp}. 
The separation was based on 
the ``RNGMOM'' cut that was placed on the quantity 
$\chi_{rm} = (R_{rs} - R_{UTC})/\sigma_R$ where 
$R_{UTC}$ ($\sigma_R$) was the expected range 
(uncertainty in range) for a given $\pi^+$ momentum and
$R_{rs}$ was the measured range in the range stack. 

\subsubsection{Delayed coincidence requirements}
\label{sec:delco}  

Determining that the incoming $K^+$ came to rest in the
target was accomplished by observing the delay between
the incoming particle and the outgoing charged track.
This requirement rejected incoming beam pions that scattered
in the target as well as the products of $K^+$ decay-in-flight. 
The delayed coincidence also served to suppress the CEX 
background.

For the standard delayed coincidence requirement, 
the average time of the kaon fiber
hits ($t_K$) had to be at least 3 ns earlier than 
the average time of the pion fiber hits ($t_{\pi}$). 
The previous \pnntwo\ analyses~\cite{ref:pnn2_PLB,ref:pnn2_PRD}  
used a tighter requirement of $t_\pi - t_K > 6\ {\rm ns}$. The 
looser requirement in this analysis resulted in a 9\% 
relative acceptance increase. 
Since under certain conditions the resolution on $t_K$ or $t_\pi$ 
was degraded, the degraded time resolution was taken into account by 
increasing 
the minimum delayed coincidence allowed.
It was increased to 4 ns 
when the energy deposit in the target kaon fibers was less than 50 MeV, 
when the time of any of the individual kaon fiber differed from the 
average kaon fiber time  by more than 2 ns 
or 
when the time of an individual target pion fiber differed from the average time of the
pion fiber hits by more than 3.5 ns.
It was increased to 5 ns 
when the difference between $t_K$ and the B4 hodoscope 
hit time was greater than 1 ns 
or 
when $t_{\pi}$ was
determined using IC hit time because the 
average time of the pion fiber hits had poor resolution. 
It was increased to 6 ns when the difference between 
$t_{\pi}$ and $t_{rs}$  was greater
than 1.5 ns. 

A tight version of the delayed coincidence with the
requirement of  $t_\pi - t_K > 6\ {\rm ns}$ 
was used to subdivide the signal region as described 
in Section~\ref{sec:like}. 
 \subsubsection{Photon veto  requirements}
\label{sec:pv_cuts}  

An event was rejected by the photon veto cut when 
the total energy in a sub-detector 
within a time window exceeded a given threshold.
The time window was referenced to $t_{rs}$, the 
reconstructed time of the pion in the range stack. 
The time window and energy threshold was 
set for each sub-detector 
using an optimization 
algorithm described in~\cite{ref:pnn1PRD}.  
The end caps were treated as three separate sub-detectors 
$EC_{outer}$, $EC_{inner}$   and $EC_{2nd}$ in the optimization.
$EC_{inner}$ was the inner ring of the 
upstream EC  and had higher accidental rates than the remaining EC elements due to its
proximity to the incoming beam. 
$EC_{2nd}$ was the EC energy identified by a double-pulse-finding algorithm 
using CCD 
information. 
$EC_{outer}$ comprised the EC elements not included in $EC_{inner}$. 
The optimization procedure 
determined  the rejection and acceptance as the time window and energy threshold were 
varied.  The optimization goal was to maximize rejection for a given value of 
acceptance. 
The acceptance sample used by the optimization procedure was derived from $K_{\mu2}$ 
monitor trigger data.

The photon veto requirements for the \pnnone\ analysis were optimized using 
 $K_{\pi2}$ peak events that
were the dominant background with photons. Ideally the \pnntwo\ photon veto
requirements would have  been optimized 
on a sample of $K_{\pi2}$ target-scatter events; however,
given that photon veto rejection needed to be ${\cal O}(2500)$,  we were
unable to prepare such a sample with 
sufficient statistics,  ${\cal O}(250000)$ events, needed to
minimize bias in the optimization result. 
In lieu of this sample, we optimized the photon veto requirements
for a majority of sub-detectors using a sample of $K_{\pi2}$ peak events 
and then optimized
the requirements for the remaining sub-detectors using 
multiple samples of $K_{\pi2}$ target-scatter events as described below.

The main sample of $K_{\pi2}$ target-scatter events 
failed either
the CCDPUL cut (Section~\ref{sec:target_cuts}) or the Beam 
Likelihood cut (Section~\ref{sec:track_cuts}) 
and contained $26317$ and $52621$ events in 
the 1/3 and 2/3 data samples, respectively.
Other $K_{\pi2}$ target-scatter samples were 
composed of events failing these cuts or the 
other target cuts described in Section~\ref{sec:target_cuts}. 
The size of the other samples ranged from $11037$ ($22037$) to 
$29899$ ($59871$) in the 1/3 (2/3) data samples. 
These samples overlapped one another, but 
they contained pions with different relative populations of
the pion scattering angle with respect to the beam direction.

An additional sample, dubbed the ``kink'' sample containing
$K_{\pi2}$ target scatters where the $\pi^+$ track 
had an identifiable kink in the \emph{xy} projection, 
was created by processing every \oneortwo\ event 
with a kinked-track reconstruction algorithm. 
For kink reconstruction, the restrictions on the pion fiber 
energy were removed as well as the requirement
that the pion fibers had to be within 1 cm of the extrapolated UTC track. 
The following criteria defined a valid kink event:
(1) the event had at least two pion fibers that  deviated from the 
UTC extrapolation, 
(2) at least 
one of the fibers from (1) must be adjacent to a kaon fiber, 
(3) the remaining pion fibers must be along the UTC extrapolation  and 
(4) the event must be rejected by the criteria placed
on the standard target-track reconstruction.  
The final criterion guaranteed that the kink sample 
was independent of the sample of signal events and 
the other samples described in the previous  paragraph. 
 Although the resulting kink sample had only 11833 events, 
it provided a sample rich in target scatters that was used 
in understanding the response of the AD as described below.

Before beginning the photon veto optimization procedure, 
we applied a cut on the activity in the BV prior
to $t_{rs}$  ($BV_{early}$) 
because 
a large energy 
deposit ($>30$ MeV) in the BV prior to the kaon decay 
would prevent the TDCs from registering activity 
coincident with $t_{rs}$~\cite{ref:ilektra}.
 The \pnnone\ set of parameters as listed in 
Table~VI of~\cite{ref:pnn1PRD} was the starting point for the \pnntwo\ optimization 
that included all sub-detectors except the AD and DPV. 
\begin{table}
\caption{\label{tab:pvparam} 
Time window and energy threshold of the primary and secondary photon veto requirements 
for each sub-detector as described in the text. 
The time window was defined with respect to $t_{rs}$, the reconstructed time of the
$\pi^+$ in the range stack. 
$RS$ and $TG$ label the range stack and target parameters, respectively. 
The parameters for the  sub-detectors below the double line were optimized separately 
as described in the text. 
$BVL_{same}$ had the additional requirement that reconstructed $z$ position satisfy $|z| < 4$ cm.
}
\begin{center}
\begin{tabular}{l c r | c r}
\hline
                        & \multicolumn{2}{c|}{Primary} 		&\multicolumn{2}{c}{Secondary}	\\
				& Time 	& Threshold	&	Time 	 	& Threshold\\
Sub-detector			&window (ns)	& (MeV)	&	window (ns)	& (MeV)	\\
\hline
$BV$			&	[-5.7,10.2]	&	0.2	&	[-7.5,10.2]	&	0.7	\\
$BVL$			&	[-4.4,10.7]	&	0.3	&	[-3.5,10.6]	&	0.3	\\
$RS$			&	[-4.3,4.4]	&	0.3	&	[-3.3,7.8]	&	0.6	\\
$EC_{outer}$		&	[-4.4,8.0]	&	0.4	&	[-6.0,9.5]	&	0.2	\\
$EC_{inner}$		&	[-3.7,5.6]	&	0.2	&	[-14.0,9.1]	&	0.2	\\
$EC_{2nd}$		&	[-5.7,2.5]	&	10.6	&	[-5.7,2.7]	&	0.2	\\
$TG$			&	[-2.7,2.2]	&	2.0	&	[-6.6,2.3]	&	1.7	\\
$IC$			&	[-2.0,4.5]	&	5.0	&	[-2.9,9.3]	&	1.4	\\
$VC$			&	[-6.6,1.8]	&	6.8	&	[-7.5,7.1]	&	5.0	\\
$CO$			&	[-0.1,5.9]	&	0.6	&	[-0.8,5.1]	&	6.0	\\
$\mu CO$		&	[-5.5,2.3]	&	3.0	&	[-4.5,3.3]	&	1.6	\\
\hline\hline
$BV_{early}$		&	[-35.7,-5.7]	&	30.0	&	[-37.5,-7.5]	&	30.0	\\
$AD$			&	[-2.0,8.0]	&	0.6	&	[-2.0,8.0]	&	0.6	\\
$DPV$			&	[-5.0,10.0]	&	0.0	&	[-5.0,10.0]	&	0.0 \\
$BVL_{same}$	        &	[-5.0,-2.0]	&	10.0	&	[-5.0,-2.0]	&	10.0	\\
\hline
\end{tabular}
\end{center}
\end{table}%
Primary and secondary sets of parameters were determined 
for the eleven sub-detectors listed in Table~\ref{tab:pvparam}.

The AD parameters were determined using the kink sample 
after application of a photon veto cut with a rejection of approximately 175 
with looser settings on the parameters of the other sub-detectors. 
These AD parameters yielded an additional photon veto rejection on the main
$K_{\pi2}$ target-scatter sample of $1.95\pm0.08$ 
with a 94\% acceptance factor~\cite{ref:ilektra}. 
The main $K_{\pi2}$ target-scatter sample was also used to optimize
the DPV parameters. 
After application of all other photon veto parameters at the primary 
setting listed in Table~\ref{tab:pvparam}, the DPV rejection was measured to be $1.13\pm0.09$ 
with an acceptance factor of 99.99\%.

The $BVL_{same}$ cut was devised 
subsequent to the single-cut failure
study on the 1/3 sample (Section~\ref{sec:singlecut}). The cut removed potential 
$K_{\pi2}$ background when both photons from the $\pi^0$ decay deposited energy 
in the same BVL element. Such an occurrence yielded a reconstructed time earlier than $t_{rs}$, 
a reconstructed $z$ position near the 
center of the element and an apparent energy greater than 10 MeV.

\begin{figure}
  \includegraphics[width=\linewidth]{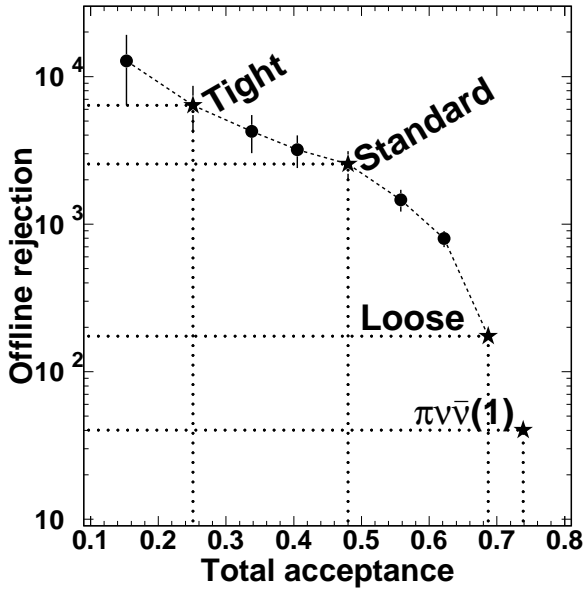}
  \caption{\label{fig:pv_rva}The offline  rejection {\sl vs.} total acceptance for the 
    optimized photon veto cuts.
    The error bars represent the statistical uncertainty.
    The labeled starred points are described in the text.}
\end{figure}
Figure~\ref{fig:pv_rva} shows the offline rejection for fixed values of 
of the total (online and offline) acceptance for the photon veto.
The parameters in the primary column in Table~\ref{tab:pvparam} corresponded to the standard
photon veto cut (``Standard'' in the Figure). 
For the more restrictive (``Tight'') photon veto cut described in Section~\ref{sec:like}, 
events were rejected that failed the criteria established
by  either the primary or the secondary parameters.
The additional settings labeled ``Loose'' and ``{\pnnone}'' in the Figure, of the photon veto cuts 
were used for background estimation (Section~\ref{sec:Muon_bkgd}) 
and consistency checks (Section~\ref{sec:otb}).

\subsubsection{Single-cut failure study}
\label{sec:singlecut}  
After determination of the selection criteria using the 1/3 data sample, 
we performed a ``single-cut'' failure study to identify unexpected 
sources of background or potential analysis flaws.
Individual cuts that exploited similar
background characteristics were grouped together to form
the following twelve cut categories:
\begin{enumerate}
        \item The cuts on $R_\pi$, $P_\pi$ and $E_\pi$ (Section~\ref{sec:kinematic_cuts}).
        \item All photon veto cuts except those on the AD and target.
        \item The photon veto cut on the AD.
        \item The target photon veto and OPSVETO (Section~\ref{sec:target_cuts}) cut.
        \item The delayed coincidence cut (Section~\ref{sec:delco}).
        \item The $\pi/\mu$ range-momentum separation 
        requirement (Section~\ref{sec:Muon_bkgd}) and the 
        pion track requirements (Section~\ref{sec:track_cuts}) excluding
        the cuts in the next two categories.
        \item {\label{item:b4ekz}} The Beam Likelihood cut (Section~\ref{sec:track_cuts}).
        \item {\label{item:tgzfool}} The cut requiring the $z$ position of the extrapolated UTC
        track to be more than 6.5 cm  from the
        upstream end of the target (Section~\ref{sec:track_cuts}).
        \item The CCDPUL and kaon fiber timing cuts (Section~\ref{sec:target_cuts}).
        \item The cuts related to the identification of the $\pi\to\mu\to e$ decay
        chain.
        \item The cuts to suppress beam-related backgrounds.
        \item The cuts on pion fiber energy, the pattern of kaon and pion fibers and the
        target-track fitter (Section~\ref{sec:target_cuts}).
\end{enumerate} 
All events in the 1/3 data sample 
that failed only one of these twelve categories were examined. 
We found four events that contained evidence
of two  potential analysis flaws.

Three of the events not rejected by the photon veto cuts 
showed evidence of a large energy deposit in the BVL. 
These events were shown to be due to $K_{\pi2}$ decays in which both
photons from the $\pi^0$ decay deposited energy in the same BVL 
counter~\cite{ref:Kentaro}. The simultaneous activity at each end 
of a BVL
element led to an erroneous average time prior to $t_{rs}$ that was
outside the  veto time window. The ``$BVL_{same}$'' cut, previously described in
Section~\ref{sec:pv_cuts}, was devised to remove these events. 

The remaining event of the four failed only the photon veto criteria in 
the AD and revealed a potential flaw in the CCDPUL target-pulse 
fitting algorithm when the fitted time of the first pulse was 
inconsistent with the average kaon fiber time. The inconsistency
arose  when the fitting algorithm incorrectly assigned the first
pulse time to an actual second pulse because the 
second pulse energy was larger than  the
first pulse energy. The CCDPUL timing criteria already described in 
Section~\ref{sec:target_cuts} were developed to remove the 
analysis flaw. 

No analysis flaws or unexpected sources of background were
revealed by the ``single-cut'' failure study of the 2/3 data 
sample.  

\subsection{Backgrounds} \label{sec:bkgd_evaluation}

\subsubsection{$K_{\pi2}$-related background}
\label{sec:Kp2g_bkgd}  

The  $K_{\pi2}$-related background contained 
 three components:
$K_{\pi2}$~target-scatter, $K_{\pi2}$~range-stack-scatter and
$K_{\pi2\gamma}$.
In order to have a $K_{\pi2}$ event
 in the signal region, 
the photons from the
$\pi^0$ decay had to escape detection
and the $\pi^+$ had to lose  energy via scattering or resolution effects
such that it fell into the signal phase space.
Scattering could happen in the target ($K_{\pi2}$ target-scatter)
or in the range stack ($K_{\pi2}$ range-stack-scatter).
The target-scatter component dominated in the background.
Since the $\pi^+$ from a $K_{\pi2\gamma}$ decay was not monochromatic,
the three final-state photons needed to escape detection for this type of
event to be a background.

\begin{figure}
  \includegraphics[width=\linewidth]{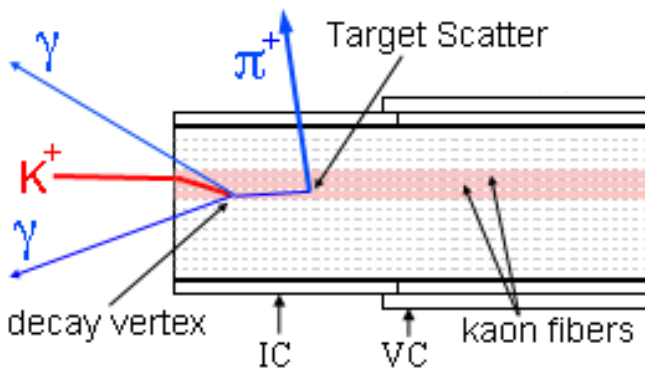}
\caption{\label{fig:scat}Schematic representation of a $K_{\pi2}$ target-scatter
in which the $\pi^+$ initially traveled in the $z$ direction, scattered
in a kaon fiber and was directed into the barrel region. The 
two photons from the decay of the recoiling $\pi^0$ were directed into the beam region.
}
\end{figure}

The topology of the most problematic type of $K_{\pi2}$ target-scatter
was that of a $\pi^+$ initially traveling in the 
kaon fibers 
and scattering into the barrel region of the detector as indicated schematically
in Figure~\ref{fig:scat}.
This type of target-scatter was
difficult to reject
because  some energy deposited in the target by the scattering $\pi^+$
occurred in a kaon fiber
(Section \ref{sec:target_cuts}) and could not always 
be
distinguished from the larger energy deposited by the stopping kaon (Figure~\ref{fig:epi_ek}). 
In addition 
the $\pi^0$ was also traveling parallel to the beam direction and
the resulting photons from the $\pi^0$ decay were directed
at the upstream or downstream ends of the detector where the
photon veto was less efficient. 

In the $K_{\pi2}$ target-scatter background estimate, the two bifurcation cuts
chosen were: the standard photon veto cuts (CUT1) and the target-quality cuts (CUT2), 
since both of these  gave powerful rejection of the 
$K_{\pi2}$ target-scatter background.
The bifurcation analysis sample was prepared by applying cuts 
to remove the contamination from muon, beam and charge-exchange events.
In the normalization branch,
a combination of $K_{\pi2}$ target-scatter, 
$K_{\pi2}$ range-stack-scatter and $K_{\pi2\gamma}$  
events were selected by inverting the photon veto cuts
($\overline{\rm{CUT1}}$). All the target-quality cuts (CUT2)
were applied to the sample, resulting in 1131 events left
in the normalization branch in the 2/3 sample. 
After corrections  for $K_{\pi2}$ range-stack-scatter
contamination (detailed below), 
$N_{\rm norm}=1107.7\pm33.8({\rm stat.})^{+2.9}_{-2.8}({\rm syst.})$ 
events remained
in the normalization branch.
The systematic uncertainty is due to the correction for range-stack-scatters.
Corrections for contamination due to $K_{\pi2\gamma}$
 are discussed later in this Section.

\begin{figure}[h!] 
  \includegraphics[width=\linewidth,height=0.35\textheight,viewport=1 5 530 530]{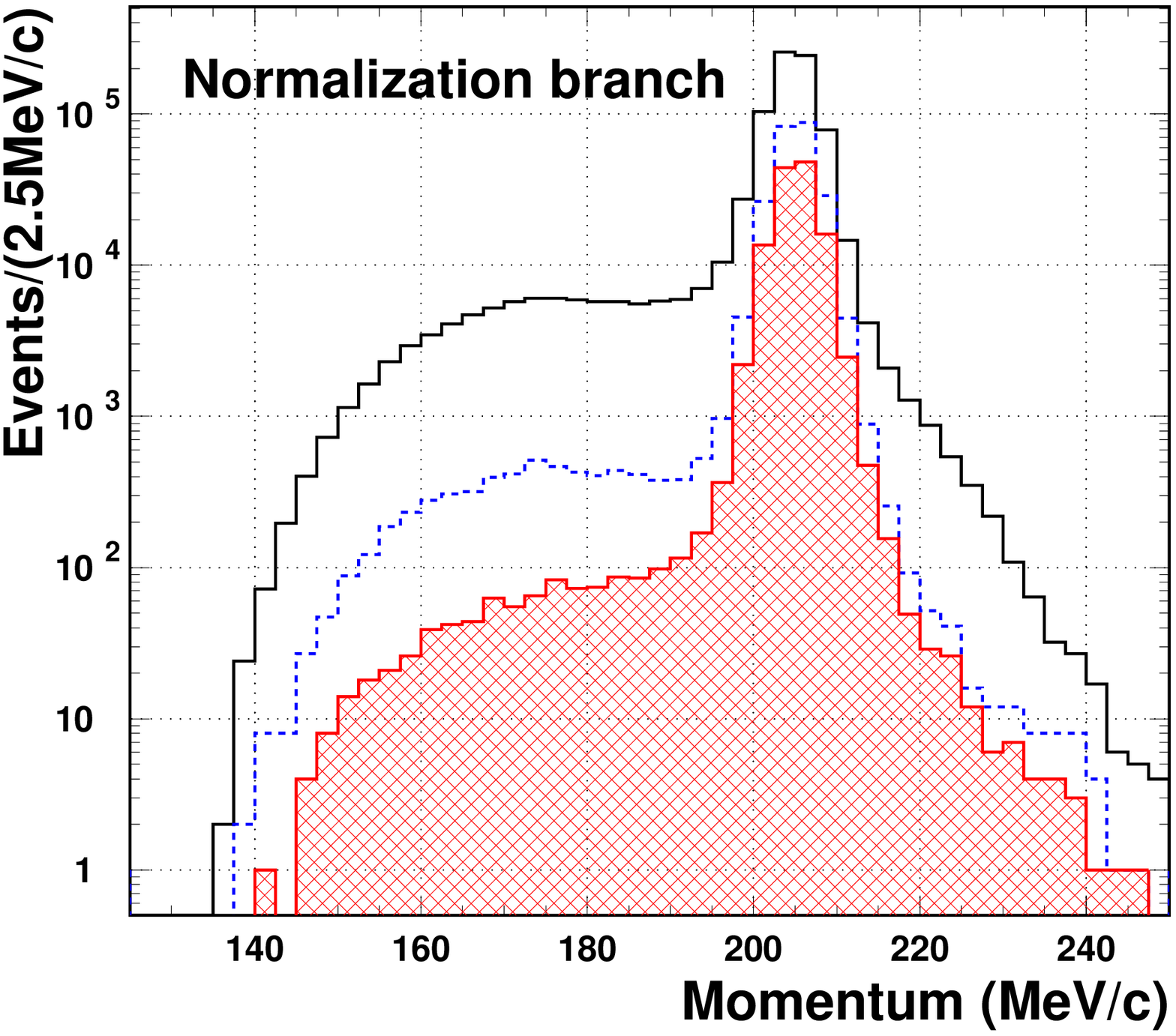}
  \includegraphics[width=\linewidth,height=0.35\textheight,viewport=1 5 530 530]{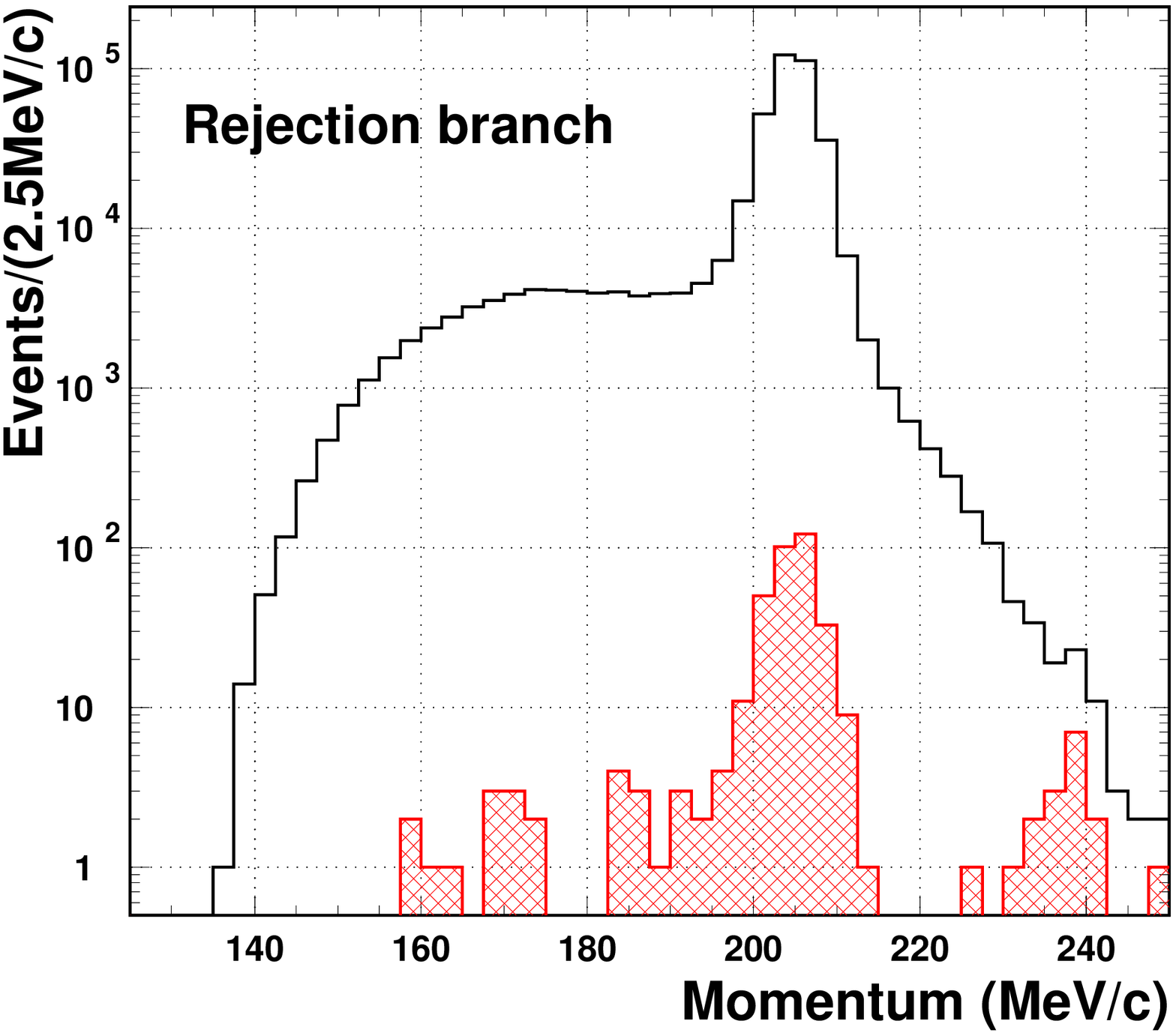}
\caption{\label{fig:kp2scat}(Top) Pion momentum distribution for the normalization branch 
in the 2/3 data sample.
The black histogram represents the distribution after inversion of the photon veto cut.
The blue dashed histogram shows the suppression of all the target cuts except for the CCDPUL cut 
and the red shaded histogram shows the suppression of all the target cuts including the CCDPUL cut.
(Bottom)  Pion momentum distribution of the rejection branch in the 2/3 data sample. 
The events obtained
from the inversion of the target cuts are shown in black and the red shaded histogram shows the events
that survive the photon veto cut. 
The events surviving the photon veto cut that peak near 236 MeV$/c$ 
were due to $K_{\mu2}$ decays. 
The cuts on pion kinetic energy and range were not applied for these distributions
in order to show the 
momentum distribution outside the \pnntwo\ signal region. 
}
\end{figure}

For the $K_{\pi2}$ target-scatter rejection branch,
the $K_{\pi2}$ target-scatter events were
classified into two non-exclusive
categories. The first category, 
``\emph{z}-scatter'', occurred when
the $\pi^+$ traveling parallel or anti-parallel to 
 the beam direction scattered in a kaon fiber 
into the barrel region of the detector as depicted in Figure~\ref{fig:scat}.
The second category, 
``\emph{xy}-scatter'',
occurred when the $\pi^+$ scattered outside of
the kaon fibers, and the scatter was visible in the \emph{xy} plane.
To measure the rejection of the photon veto for target-scatter events,
six classes of
$K_{\pi2}$ target-scatter events containing varying mixtures of \emph{xy}-scatter
and \emph{z}-scatter events
were created by applying or inverting
various combinations of the requirements on $\pi^+$ in the target
(Section \ref{sec:target_cuts}).
The primary $K_{\pi2}$ target-scatter sample (Section~\ref{sec:pv_cuts}), 
considered to be the richest in \emph{z}-scatters, 
was chosen to measure the photon veto rejection, giving 52621 events for the region 
$C+D$ (Figure~\ref{fig:bifurc}).
The photon veto cuts (CUT1) were then applied to the remaining 
$K_{\pi2}$ target-scatter events, leaving 22 events for the region $C$ 
for a rejection of $52621/22 = 2392\pm510$, where the uncertainty is statistical only. 
The pion momentum distributions of
the normalization and rejection branches are shown in Figure~\ref{fig:kp2scat}. 
The photon veto rejections measured in the other five classes of target-scatter events 
were consistent with that of the primary sample and extended 
from a minimum of $2204\pm697$ to a maximum of $2758\pm650$. 
The photon veto rejection of the main $K_{\pi2}$ target-scatter
sample in the tighter kinematic region (Section~\ref{sec:kinematic_cuts}) 
was measured to be $2193\pm517$ in agreement with the measurement
in the standard kinematic region. 
The range of measured rejection values in the five 
other classes was used to set the systematic
uncertainty in the photon veto rejection on the $K_{\pi2}$ target-scatters,
giving a photon veto rejection of 
$2392\pm510^{+366}_{-188}$. 
Using Equation~(\ref{eqn:bif}), 
the uncorrected number of $K_{\pi2}$ target-scatter background events was measured to be
\begin{eqnarray}
b^{tg}_{\rm un}
 & =&  3/2 \times(1107.7\pm33.8^{+2.9}_{-2.8})/( (2392\pm510^{+366}_{-188})-1) \nonumber \\
 & =& 0.695\pm0.150^{+0.067}_{-0.100},
\end{eqnarray} 
where the first uncertainty was statistical and the second uncertainty
systematic. 

For the $K_{\pi2}$ range-stack-scatter background events, 
the cuts with the most powerful rejection were
the range-stack track quality  and the photon veto cuts.
The $K_{\pi2}$ range-stack-scatter normalization branch
was a modified version of the 
$K_{\pi2}$ target-scatter normalization branch,
with the range-stack quality cuts inverted instead of being applied
before the inversion of the photon veto cut as was done in the 
$K_{\pi2}$ target-scatter normalization branch.
This sample of $N_2=281$ events was heavily contaminated 
with target-scatter events due to the
inefficiency of the range-stack-scatter cuts.
The $N_1=1131$ events remaining at the end of the $K_{\pi2}$ target-scatter
normalization branch consisted of $N^{tg}$
target-scatter events with contamination due to $N^{rs}$
range-stack-scatter events:
\begin{equation}\label{eqn:kp2norm}
N^{tg} + N^{rs} = N_1.
\label{eqn:tgscat}
\end{equation}
These $N^{tg}$
$K_{\pi2}$ target-scatter events and $N^{rs}$
$K_{\pi2}$ range-stack-scatter events were also related to the $N_2$
 events remaining in the 
$K_{\pi2}$ range-stack-scatter normalization branch by
\begin{equation}
\frac{1-A^{rs}}{A^{rs}}\times N^{tg} 
+ \left( R^{rs}-1 \right)\times N^{rs} = N_2,
\label{eqn:rsscat}
\end{equation}
where $A^{rs}=0.888\pm0.001({\rm stat.})\pm0.012({\rm syst.})$
was the acceptance factor for  the range-stack quality cuts
and $R^{rs}=7.06\pm0.47$ 
was the rejection of $K_{\pi2}$ range-stack-scatter
events by the range-stack quality cuts 
measured using events with momentum consistent with 
 the $K_{\pi2}$-peak region, but range and energy in the \pnntwo\   signal region 
as would be expected for a range-stack-scatter. 
The systematic uncertainty on $A^{rs}$ was due to the larger uncertainty on
the measured kinematics of the $\pi_{\rm{scat}}$ monitor data used
to measure the acceptance of the range-stack quality cuts as described in 
Section \ref{sec:acceptance}.
By solving Equations~(\ref{eqn:tgscat}) and (\ref{eqn:rsscat}) simultaneously,
it was possible to estimate the number of $K_{\pi2}$ target-scatter events
($N^{tg}=1107.7\pm33.8^{+2.9}_{-2.8}$)
and the number of $K_{\pi2}$ range-stack-scatter
events ($N^{rs}=23.3\pm3.5^{+2.9}_{-3.0}$) present 
in the original $K_{\pi2}$ target-scatter normalization branch,
where the first uncertainty is statistical and
the second is systematic due to the acceptance factor $A^{rs}$. 

The photon veto rejection on the $K_{\pi2}$ range-stack-scatter events should be 
the same as that for the unscattered $K_{\pi2}$ peak events as the 
back-to-back correlation of the $\pi^+$ and $\pi^0$ was maintained. 
The $K_{\pi2}$ range-stack-scatter rejection branch was created
by applying all  cuts other than the photon veto cuts and the
 pion kinematic  cuts (Section~\ref{sec:kinematic_cuts}).
The
$K_{\pi2}$ peak region events were selected, creating a sample
of 122581 events for the region $C+D$.
The PV cuts (CUT1) were then applied to the remaining 
$K_{\pi2}$ events, leaving 106 events in region $C$ for a 
photon veto rejection of $122581/106=1156\pm112$.
The number of $K_{\pi2}$ range-stack-scatter background events was measured to be
\begin{eqnarray}
b^{rs} &=& 3/2 \times(23.3\pm3.5^{+2.9}_{-3.0})/( (1156\pm112) -1) \nonumber\\
       &=& 0.030\pm0.005({\rm stat.})\pm0.004({\rm syst.}) .
\end{eqnarray}

The $K_{\pi2\gamma}$ background estimate did not use the bifurcation method,
but used  a combination
of $K_{\pi2}$ events selected in the \oneortwo\ data and simulated $K_{\pi2}$ and $K_{\pi2\gamma}$ events. 
We simulated both the inner bremsstrahlung (dominant) and direct emission amplitudes
of $K_{\pi2\gamma}$ decays assuming no interference between them~\cite{ref:NA482}.
The $K_{\pi2\gamma}$ background was estimated as
\begin{equation}
b^\gamma = \frac{N}{\kappa R_\gamma}
\end{equation}
where
\begin{itemize}
\item[$N$]$=106$ was the number of $K_{\pi2}$-peak  events in the \oneortwo\ trigger
  sample after all selection criteria were applied except that the $K_{\pi2}$-peak region
  kinematic region was selected.
\item[$\kappa$]$=417\pm24$ was the ratio of acceptance factors of $K_{\pi2}$ to $K_{\pi2\gamma}$
  events as determined from simulated $K_{\pi2}$ and $K_{\pi2\gamma}$ decays 
  taking into account the $K_{\pi2}$ and $K_{\pi2\gamma}$ branching ratios.
  We used ${\cal B}(K^+\to\pi^+\pi^0\gamma) = (11.1\pm0.6)\times 10^{-4}$ obtained
  by correcting the partial branching ratio 
  $(2.75\pm0.15)\times10^{-4}$~\cite{ref:PDG} measured for $55<P_\pi<90\ {\rm MeV}$ 
  to the full energy range using simulation.
\item[$R_\gamma$]$=5.04\pm0.10$  was the additional photon veto rejection afforded by the
  radiative photon. This additional rejection 
  was calculated by combining the 
  distribution of the radiative photon from simulated events 
  with the measured single photon detection efficiency as a function of 
  angle and energy from $K_{\pi2}$ data \cite{ref:Kentaro}.
\end{itemize}
The final anticipated number of $K_{\pi2\gamma}$ background events 
was 
$b^\gamma =  0.076\pm0.007\pm0.006$,
where the first uncertainty was statistical  and the
second was systematic (due to $\kappa$ and $R_\gamma$).

The inverted photon veto used to select events for the
$K_{\pi2}$ target-scatter normalization branch would have 
also selected  $K_{\pi2\gamma}$ events.
We corrected 
the estimate for the $K_{\pi2}$ target-scatter background
by subtracting $b^\gamma$, 
\begin{eqnarray}
b^{tg} & = & b^{tg}_{\rm un} - b^\gamma \nonumber \\
      & = & 0.619\pm0.150^{+0.067}_{-0.100} \ \ .
\end{eqnarray}

\subsubsection{ $K^+ \rightarrow \pi^+ \pi^- e^+ \nu$ background}
\label{sec:Ke4_bkgd}  

Despite the small  branching ratio
of $(4.09\pm0.10)\times10^{-5}$~\cite{ref:PDG}, $K_{e4}$ 
could be a background if the  $\pi^-$ and the $e^+$
escaped detection in the target. 
\begin{figure}
\begin{center}
\includegraphics[width=\linewidth]{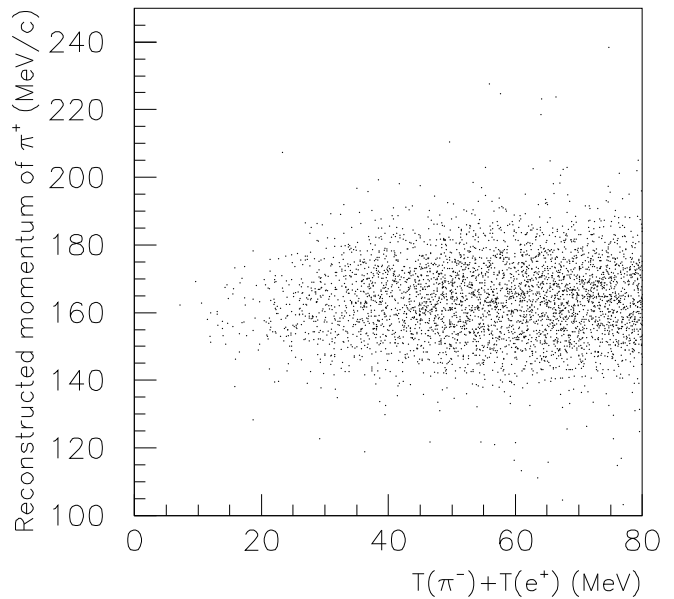}
\caption{\label{fig:ke4_t2}The sum of the $\pi^-$ and $e^+$ kinetic 
energies versus the reconstructed
momentum of the $\pi^+$ for simulated $K_{e4}$ events that passed the \oneortwo\ trigger.}
\end{center}
\end{figure}
The distribution of the sum of the kinetic energies  of the $\pi^-$ and the $e^+$ 
{\sl vs.} the reconstructed $\pi^+$ momentum in simulated
events  shown in Figure~\ref{fig:ke4_t2} indicates
where the $K_{e4}$ background would occur kinematically.

\begin{figure} 
\begin{center}
\includegraphics[width=\linewidth]{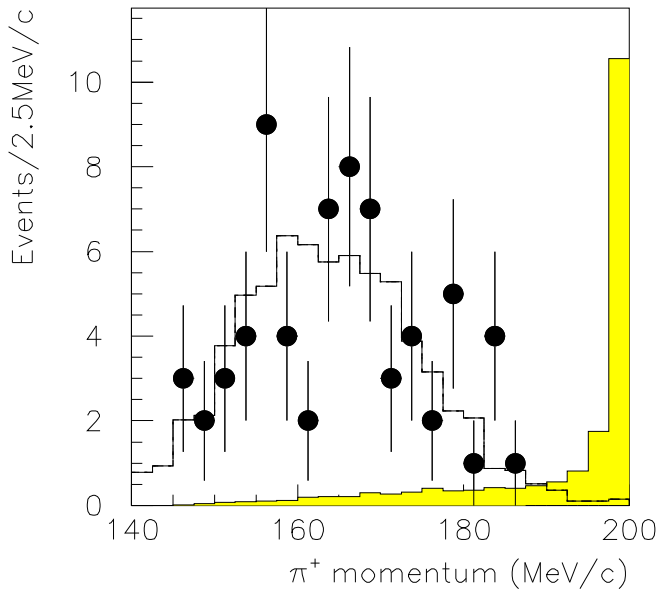}   
\caption{\label{fig:ke4_norm}Momentum distribution of the $\pi^+$ in the $K_{e4}$ normalization branch 
for the 1/3 sample before the application of the CCDPUL cut. The points 
 represent the data. The unshaded histogram is the distribution as expected from 
simulated $K_{e4}$ events. The histogram area is normalized to the number of data events.
The shaded histogram represents the 
normalization branch for $K_{\pi2}$ target-scatter events.
The ratio of the area of the shaded to the unshaded histogram has been arbitrarily set to 
1/8 times the ratio of the  $K_{\pi2}$ target-scatter background to the $K_{e4}$ 
background for display purposes.}
\end{center}
\end{figure}

Since the  main characteristic of $K_{e4}$ event was extra energy
in the target from the $\pi^-$ and the $e^+$, the target photon veto (TGPV), 
OPSVETO and CCDPUL cuts were the most effective cuts to suppress this background.
Due to contamination by other types of background, such as $K_{\pi2}$-target-scatter, 
it was not possible to isolate a pure $K_{e4}$ background sample for a 
 bifurcation analysis using data only. Nonetheless, a $K_{e4}$-rich sample was selected from 
data using the  $\rm CCDPUL\cdot\overline{\rm TGPV\cdot OPSVETO}$ requirement and 
served as the normalization branch.
We established that the majority of the events in the normalization branch were
likely to be due to $K_{e4}$ decays by removing the CCDPUL requirement and 
comparing the momentum distribution of the selected events in the 1/3 sample with the
expectation from simulation (Figure~\ref{fig:ke4_norm}). 
In addition, the target information for the 69 events in Figure~\ref{fig:ke4_norm} 
was visually examined and the events classified based on the topology, ionization pattern, 
curvature, range and energy of the putative tracks. 
Fifty-nine events were classified as $K_{e4}$, three as $K^+\to\pi^+\mu^-\mu^+$ 
and four as $K_{\pi2}$ or $K_{\pi2\gamma}$ (including events with apparent $\pi^0\to e^+e^-\gamma$ decays).
The classification of the remaining three events was ambiguous. Assuming half the ambiguous
events were $K_{e4}$ yielded a purity of $\sim\!88\%$.
One 
example of a $K_{e4}$ candidate event is shown in Figure~\ref{fig:target1}.

Simulated $K_{e4}$ events were used to determine the rejection of the  TGPV, OPSVETO, and 
CCDPUL requirements. 
Negative pion absorption in the target was modeled based on the 
energy spectrum of stopped $\pi^-$ in plastic scintillator 
observed in E787~\cite{ref:piminus}. We 
assumed that all energy generated from $\pi^-$ absorption would be
promptly deposited in the single fiber where the $\pi^-$ came to rest. 
This assumption conservatively neglected the possibility that 
detectable activity from $\pi^-$ absorption could occur elsewhere in the 
detector~$\!\!$\protect\footnote{Negative pions are predominantly absorbed by 
carbon nuclei in
scintillator resulting in multi-nucleon emission. 
The measured~\cite{Hufner:1975ys,PhysRevC.19.135} 
rates of emission per stopped $\pi^-$ of neutrons,
 protons, deuterons, tritons and alphas are 
approximately 
2.8, 0.3, 0.2, 0.1 and 0.6, respectively, with typical 
kinetic energies of tens of MeV.
Because of their short range, 
these charged particles will deposit energy very 
close to their absorption points.
The mean free path of emitted neutrons is tens of cm leading to energy deposition 
relatively far from the absorption point. In addition, the residual nucleus 
is unstable and 
can deexcite by emission of photons with typical energies of  1--2 MeV.
The energy spectrum measurement~\protect\cite{ref:piminus} was sensitive to
energy near the absorption point and thus largely neglected any additional energy
deposition due to the latter two processes.}.
Positron interactions were well-modeled in 
our EGS4-based simulation~\cite{ref:EGS}. 
The rejection of the CCDPUL, TGPV and OPSVETO requirements were correlated
because the  target fibers containing the deposited energy of the $\pi^-$ and $e^+$ 
could have been classified as kaon, pion, photon or opposite-side pion fibers.
We used the energy of the simulated deposits to estimate the rejection 
of these cuts as $R=52^{+121}_{-29}$. 
As we did not precisely model either $\pi^-$ absorption or 
the inactive material of 
the target such as the gaps between the fibers and the cladding and wrapping 
material of each fiber, we varied the threshold for the energy treated by the CCDPUL 
($\rm TGPV\cdot OPSVETO$) cut by a factor of 5 (1.5) to estimate the systematic uncertainty
associated with the rejection of these cuts.
The normalization branch in the 2/3 sample contained 6 events so the
 $K_{e4}$ background was measured to be 
$3/2 \times 6/(52^{+121}_{-29}-1)  = 0.176\pm0.072^{+0.233}_{-0.124}$ 
events, where the first error was statistical and the second was systematic.

\subsubsection{Muon background}
\label{sec:Muon_bkgd}  

The decays $K^+\to\mu^+\nu_\mu$,  
$K^+ \rightarrow \mu^+\nu_\mu\gamma$ and $K^+ \rightarrow \mu^+\pi^0\nu_\mu$ could
contribute background in the \pnntwo\ kinematic region 
as indicated in Figure~\ref{fig:triggers_rp}.
All three processes  required the muon 
to be mis-identified as a pion in order to 
be a background. 
The first
decay would be background if the kinematics 
of the $\mu^+$ were mis-reconstructed and
the latter two decays would be background if 
the photons went undetected.

The two bifurcation cuts were  
the  $\pi\to\mu\to e $ identification or ``TD'' cut (CUT1),  
and the $\pi/\mu$ range-momentum separation or ``RNGMOM''  (CUT2). 
The normalization branch defined by inverting the TD cut yielded zero events in 
the 2/3 sample, so $N_{\rm norm}$ was assigned to be 1 event.

The $\mu^+$ rejection branch  
contained  $C+D=20488$ events in the 2/3 sample
and 
was selected by inverting CUT2 and 
applying cuts to remove beam backgrounds 
and the \pnnone\ version of the photon veto cut (Figure~\ref{fig:pv_rva}) 
to suppress $K_{\pi2}$ backgrounds.  
 After the application of the 
TD cut, the number of  events remaining was $C=154$ 
for a  measured  TD cut rejection of $133.0\pm10.7$.
Thus, the $\mu^+$ background was estimated  to be 
$3/2\times (1\pm1)/((133.0\pm10.7)-1) = 0.0114\pm0.0114$.

\subsubsection{Charge-exchange background}
\label{sec:CEX_bkgd}  

When the $K^0$ due to CEX in the target 
decayed as a $K^0_L$ it was a potential background.
The delayed coincidence requirement effectively removed any contribution
from the short-lived $K^0_S$. The semileptonic decay processes
$K^0_L\rightarrow \pi^+e^-\bar\nu_e$ and $K^0_L\rightarrow \pi^+\mu^-\bar\nu_{\mu}$
with branching ratios of 20\% and 14\%, respectively, were 
considered to be the most likely
to form a background.

The CEX background could also contain a component 
due to hyperon production 
where a $\pi^+$ was either produced with the hyperon or was 
a hyperon decay product. 
Hyperon production would result from $\overline{K^0}$-nucleon interactions if
the $K^0$ oscillated to a $\overline{K^0}$.

\begin{figure}
\begin{center} 
\includegraphics[width=0.8\linewidth,height=0.20\textheight]{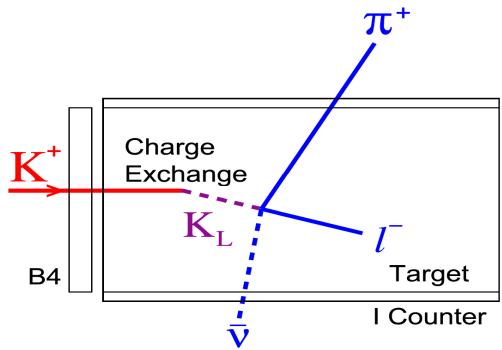}
\caption{\label{fig:CEX}Schematic diagram of the charge-exchange process 
  in the target.}
\end{center}
\end{figure}

Simulation studies showed that there was often a gap between the pion and 
kaon fibers
and that the reconstructed $z$ of the pion track was not consistent with 
the energy deposited 
in the kaon fibers  
as indicated schematically in Figure~\ref{fig:CEX}.
A CEX-rich sample that served as the normalization branch was selected 
in \oneortwo\ triggers 
by requiring a gap between the pion and kaon fibers. 
No target energy cuts were applied in the selection of the normalization 
sample since
the lepton from $K^0_L$ decay or the $\pi^-$ and $\pi^0$
associated with hyperon production might deposit extra energy in the target.
The offline delayed coincidence requirement was also not applied for the 
normalization 
sample. In contrast to the previous \pnntwo\ analysis~\cite{ref:pnn2_PLB}, 
the ability 
to create a normalization sample from \oneortwo\ data avoided uncertainties
 associated
with the effective CEX cross-section and related efficiencies. 
In addition a normalization sample selected from the data contains contributions 
from all CEX processes including $K^0_L$ and hyperon decays. 
The normalization branch in the 2/3 sample contained one event.

The rejection associated with finding a gap in the CEX events, 
the target energy cuts  and the delayed coincidence
was determined from simulated CEX $K_L^0\to\pi^+\ell^-\bar\nu$ events.
For the simulation of CEX events, 
reconstructed  $K^0_S \rightarrow \pi^+ \pi^-$ events
obtained from the CEX monitor trigger data (Section~\ref{sec:trigger})
were used as the $K^0_L$ production point and momentum. 
The rejection of the delayed coincidence and gap-finding cuts exploit the
flight of the neutral kaon and should be the same for $K_L^0\to\pi^+\ell^-\bar\nu$ and
hyperons. The rejection of the target energy cuts on the $\pi^-$ or $\pi^0$ associated 
with hyperon decay was estimated to be comparable to or greater than the rejection of 
the target energy cuts  on the lepton from $K^0_L\to\pi^+\ell^-\bar\nu$ decays.
The background was measured in the 2/3 sample 
to be $0.013\pm{0.013}({\rm stat.})^{+0.010}_{-0.003}({\rm syst.})$ events. 
The systematic uncertainty was estimated by the same threshold variations 
 of the target energy cuts that were used  for $K_{e4}$ (Section~\ref{sec:Ke4_bkgd}).


\subsubsection{Single-beam background} \label{sec:singlebeam}

The bifurcation cuts for single-beam background were the 
delayed coincidence (CUT1) and B4 energy 
of less than 1.0 MeV (CUT2). CUT2  selected beam pions.  
The sample was selected by applying all the photon veto cuts except TGPV, 
the kinematic cuts, TD cuts and  beam cuts except the delayed coincidence 
and CUT2.  
The normalization sample formed by the inversion of the 
 delayed coincidence cut 
samples  yielded zero events, so that $N_{\rm norm}$ was assigned 
to be $1$.
The  rejection sample contained $C+D=12850$ events in the 2/3 data. After 
the application of the delayed coincidence,  $C=2$ events remained
for a rejection of the delayed coincidence cut of $6425\pm4543$.  
 The measured single-beam background was 
$3/2\times(1\pm1)/(6425-1) = 0.00023\pm0.00023$ events.  

\subsubsection{Double-beam background} \label{sec:doublebeam}

Double-beam background had two components, $KK$ and $K\pi$. For the $KK$ ($K\pi$) 
background, the decay products of the initial kaon were undetected and a subsequent
kaon decay  (a scattered beam pion) provided the outgoing $\pi^+$. 

The $KK$ component of the  background $b^{KK}$ was determined using the 
bifurcation procedure described in Section~\ref{sec:analmeth} with modifications
to compensate for poor statistics:
\begin{equation}\label{eqn:bkk}
b^{KK}
=
f
\times 
\frac{n_{KK}/r_{KK}}{R_{KK}-1}
\end{equation}
where $f=3/2$ was the scale factor for the 2/3 data sample (Equation~(\ref{eqn:bif})). 
\begin{itemize}
        \item[$R_{KK}$]$=1576/4 = 394\pm197$ was the measured rejection of the kaon
        \v{C}erenkov and BWPC cuts (CUT1) on a beam kaon at the time
        of the outgoing pion ($t_{rs}$). The rejection sample was prepared
        by vetoing beam  pions at $t_{rs}$ via \v{C}erenkov information,
        by requiring a second track at $t_{rs}$ in the B4 counter with an energy
        deposit consistent with a kaon and by  the more stringent target gap
        requirement described in Section~\ref{sec:target_cuts}. The latter criterion
        ensured activity in the target in two spatially and temporally distinct
        regions indicative of a double-beam event.
        \item[$\frac{n_{KK}}{r_{KK}}$] was the normalization provided by a second bifurcation
        of the standard normalization branch. The second bifurcation~\cite{ref:pnn1PRD} exploited the
        lack of correlation between the AD and target cuts to improve the statistical
        power of the measurement. The normalization branch was prepared by inverting CUT1,
        by vetoing entering pions at $t_{rs}$ using \v{C}erenkov information and 
        by application of the $\pi\to\mu\to e$ and track quality requirements on the
        outgoing track and contained 2699 events. The application of the AD photon
        veto cuts reduced the sample to 325 events for a rejection of $r_{KK}=8.3\pm0.4$.
        The application of the target cuts on the 2699 events yielded zero events
        so we assigned $n_{KK} = 1$.
\end{itemize}
These values gave the $KK$ background of 
$b^{KK}   =  0.00046\pm0.00046$ events.

An analogous method was used to estimate the $K\pi$ component of the double-beam
background
\begin{equation}\label{eqn:bkp}
b^{K\pi}
=
\frac{1}{1 - 0.606}
\times
f 
\times
\frac{n_{K\pi}/r_{K\pi}}{R_{K\pi}-1}
\end{equation}
where the additional scale factor of ${1}/(1 - 0.606)$
was included to correct for the data
accumulated with the online pion \v{C}erenkov veto in the \pnntwo\ trigger 
(Section~\ref{sec:trigger}). With the online veto, the offline rejection of the
pion \v{C}erenkov cuts was low and the normalization branch lacked statistics. 
 To obtain $b^{K\pi}$, the $K\pi$ background estimate 
obtained  without the online veto was scaled by the ratio of the kaon exposures. 
As a check of this scaling,
we verified that the $KK$ background estimates for the two trigger configurations
were consistent and that the $K\pi$ background for the \pnnone\ trigger data
was consistent for the two data-taking periods~\cite{ref:benji}. With the measured values 
of $R_{K\pi} = 2467/4 = 617\pm308$, $n_{K\pi} = 1\pm1$ and $r_{K\pi} = 4435/464 = 9.6\pm0.4$ 
in Equation~(\ref{eqn:bkp}), 
we obtained $b^{K\pi}= 0.00064\pm0.00064$ events.

\subsubsection{Background summary}
\label{sec:bkgd_summary}

The contribution of each background component is 
listed in Table~\ref{tab:bkgd}.
\begin{table}[h] 
\begin{center}
\begin{tabular}{l|l}
\hline
Process           & Background events       \\ \hline
\KPtwo\ target-scatter      &$0.619\pm0.150{}^{+0.067}_{-0.100}$      \\
\KPtwo\ range-stack-scatter      &$0.030\pm0.005\pm0.004$                \\
$K_{\pi2\gamma}$   &$0.076\pm0.007\pm0.006$               \\
$K_{e4}$          &$0.176\pm0.072{}^{+0.233}_{-0.124}$     \\
Muon               &$0.011\pm0.011$                     \\
Charge-exchange                &$0.013\pm{0.013}{}^{+0.010}_{-0.003}$ \\
Beam               &$0.001\pm0.001$                     \\ \hline
Total              &\TOTbkgd\                           \\ \hline
\end{tabular}
\caption{\label{tab:bkgd}
Summary of the estimated number of events in the signal region 
from each background component. 
Each component is described in the text.}
\end{center}
\end{table}
The total background was estimated to be \TOTbkgd\ events and was
dominated by the  $K_{\pi2}$ target-scatter component that 
was the largest contribution to the statistical uncertainty.
The systematic uncertainty was dominated by the contribution from 
the $K_{e4}$ background due to the inability to establish a 
precise correspondence between the energy observed in the target
in data and simulation. 

A number of background consistency and validity checks were
performed  as described below.  
\subsubsection{Background contamination evaluation}
\label{sec:contamination}  
Due to the difficulty of isolating background samples, 
studies were performed to estimate
the degree of contamination 
({\sl i.e.}, events due to background from other sources) 
in the $K_{\pi2}$ target-scatter
normalization and rejection branches (Figure~\ref{fig:kp2scat}).

The effect of muon contamination of the $K_{\pi2}$ background estimate
was determined separately for the normalization and rejection branches
with and without the $\pi\to\mu\to e$ (TD)  and $\pi/\mu$ range-momentum 
separation (RNGMOM) cuts. The normalization branch used in the 
$K_{\pi2}$-scatter study was assumed 
to be the sum of $\pi^+$ and $\mu^+$ components such that
\begin{equation}\label{eqn:contam1}
N_{\rm norm} = 1131 = N^\pi_{\rm norm} + N^\mu_{\rm norm}  \ \ .
\end{equation}
When the TD and RNGMOM cuts were not applied, the observed number
of events in the normalization branch was
\begin{equation}\label{eqn:contam2}
n = 12980 = N^\pi_{\rm norm}/A_\pi + R_\mu N^\mu_{\rm norm}
\end{equation}
where $A_\pi$ ($R_\mu$) is the acceptance (rejection) factor 
for the combination of the TD and RNGMOM cuts for pions (muons).
The rejection of the TD cut was evaluated as $133.0\pm10.7$ as part of the
muon background estimate (Section~\ref{sec:Muon_bkgd}). The RNGMOM
rejection of $28.3\pm1.1$ was evaluated using the muon normalization
branch for a total muon rejection of $R_\mu = 3764\pm333$. 
The acceptance factor $A_\pi$ of the combination of the TD and
RNGMOM cuts was determined on samples of $K_{\pi2}$-peak region events
that failed different combinations of the target-scatter cuts 
used to assess the uncertainty in the photon veto rejection as 
described in Section~\ref{sec:pv_cuts}. The acceptance factors
for these samples, both before and after the application of the
standard photon veto cut, were consistent and yielded
$A_\pi = 0.809\pm0.030$. These gave the muon contamination of the
normalization sample of $N^\mu_{\rm norm}/N_{\rm norm} = (2.7\pm0.3)\times10^{-3}$. 
Analogous methodology was used to assess the effect of muon 
contamination in the rejection branch. The calculated photon veto
rejection after correction for muon contamination was $R^\pi = 2410\pm518$ to be 
compared with $R=2392\pm510$ (Section~\ref{sec:Kp2g_bkgd}). 
Inserting these results into the background estimate using the bifurcation 
method (Equation~(\ref{eqn:bif}))  
implied that the muon contamination increased the $K_{\pi2}$ background ($b$) 
estimate by 
\begin{eqnarray*}
b/b^\pi &=& f\frac{N_{\rm norm}}{R-1} \Big/ f\frac{N^\pi_{\rm norm}}{R^\pi-1} \nonumber \\
      &=& 1.010\pm0.002 
\end{eqnarray*}
which was considered negligible with respect to the estimated systematic
uncertainty. 

A similar treatment limited the overestimate of the $K_{\pi2}$ background
due to double-beam contamination to be $<0.1\%$.


The rejection of $K_{e4}$ by the photon veto should be less than
the photon veto rejection of $K_{\pi2}$ and $K_{\pi2\gamma}$ in that there were
no photons in the final state. Contamination of the $K_{\pi2}$ rejection
sample by $K_{e4}$ events would therefore reduce the measured 
photon veto rejection.
The measured rejection of $39481/18 = 2193\pm517$
in the tighter kinematic region, 
that was defined to suppress $K_{e4}$ (Section~\ref{sec:kinematic_cuts}), 
was consistent with the overall rejection of $2392\pm510$ 
(Section~\ref{sec:Kp2g_bkgd}),  
indicating no significant contamination of the $K_{\pi2}$ rejection sample
by $K_{e4}$. 
The $K_{\pi2}$ normalization branch was defined by the inversion of the
photon veto cuts and the application of the target-quality cuts, including
CCDPUL and OPSVETO. The $K_{e4}$ normalization was prepared by
application of $\rm CCDPUL \cdot\overline{\rm TGPV\cdot\rm OPSVETO}$. 
Since $\overline{\rm TGPV}\cdot\rm OPSVETO\cdot\rm CCDPUL$ was a
subset of the $K_{\pi2}$ normalization branch,  the
contamination of the $K_{\pi2}$ normalization branch by $K_{e4}$ was
less than the six events selected in the $K_{e4}$ normalization 
branch (Section~\ref{sec:Ke4_bkgd}) and hence negligible compared
to the 1131 events in the $K_{\pi2}$ normalization branch (Equation~(\ref{eqn:kp2norm})).

\subsubsection{Background consistency checks}
\label{sec:otb}  

The consistency of the background estimate was checked in 
three distinct data regions just outside the signal region that were 
created by loosening the photon veto cut  and the CCDPUL cut.
The region $CCD_1$
was immediately adjacent to the signal region and contained 
events with a CCDPUL second-pulse energy above the 
standard threshold of 1.25 MeV and below 2.5 MeV. 
The region $PV_1$ was  immediately adjacent to the signal region
and 
defined by events rejected by the standard photon veto and
accepted by the loose photon veto cuts (Figure~\ref{fig:pv_rva}).
The region $PV_2$ was  adjacent to $PV_1$ 
and 
defined by events rejected by the loose photon veto and 
accepted by the \pnnone\ photon veto cuts (Figure~\ref{fig:pv_rva}).
The numbers of expected background events in these regions 
were calculated in the same manner as for the signal region.

Table \ref{tab:otb} shows the number of expected and
observed events in the three regions as well as the probability 
of the observed number of events or fewer given the expectation.
The combined probability of 5\% for the two regions
nearest the signal region may have indicated that the background 
was overestimated, but the re-evaluation of this combined probability
at the lower limit of the systematic uncertainties \cite{ref:footnote} gave
14\% for the two closest regions which demonstrated that
the assigned systematic uncertainties were reasonable.
\begin{table}[h]
\begin{tabular}{c|c|c|c|c}
\hline
Region & $N_E$ & $N_O$ & ${\cal{P}}(N_O;N_E)$ & Combined \\
\hline
$CCD_1$ & $0.79^{+0.46}_{-0.51}$ & 0 & 0.45 [0.29,0.62] & \\[0.3mm] 
$PV_1$ & $9.09^{+1.53}_{-1.32}$ & 3 & 0.02 [0.01,0.05] & 0.05 [0.02,0.14] \\ [0.3mm] 
$PV_2$ & $32.4^{+12.3}_{-8.1}$ & 34 & 0.61 [0.05,0.98] & 0.14 [0.01,0.40] \\
\hline
\end{tabular}
\caption{\label{tab:otb}
 Comparison of the expected ($N_E$) and observed ($N_O$) number of background events
in the  three regions $CCD_1$, $PV_1$, and $PV_2$ outside the signal region.
 The central value of $N_E$ is given along with the combined statistical and 
 systematic uncertainties. ${\cal{P}}(N_O;N_E)$ is the probability of observing
 $N_O$ or fewer events when $N_E$ events are expected. The rightmost column
 ``Combined'' gives the probability of the combined observation in that region and the
 region(s) of the preceding row(s). The numbers in square brackets are the 
 probabilities reevaluated at the upper and lower bounds of the uncertainty 
on $N_E$~\cite{ref:footnote}. } 
\end{table}

The assignment of $N_{\rm norm}=1$ when no events were observed in the
normalization branch (Section~\ref{sec:analmeth}) was only made for the
muon, single-beam and $KK$ double-beam backgrounds. Thus, this assignment 
could have overestimated the total background by, at most, 0.012 events or 1.3\%.

\subsection{Acceptance and sensitivity} \label{sec:acceptance}

We assessed the overall acceptance of all selection criteria by 
dividing the criteria into components that could be measured 
separately using monitor triggers or simulated data. 
Simulated data were used to estimate the acceptance of the 
trigger and decay phase space as well as to assess the
impact of nuclear interactions. 
The overall acceptance was the product of the acceptance factors
for each component. Correlated cuts were grouped together for evaluation.

\subsubsection{Acceptance factors from  $K_{\mu2}$ events}

$K_{\mu2}$ monitor triggers were used to assess the components of
the acceptance regarding the kaon beam, the charged track, 
the event topology and the standard photon veto. The acceptance factors are listed in
Table~\ref{tab:km2acc} and described below.
\begin{table}[h]
\caption{\label{tab:km2acc}Acceptance factors of the \KPpnn\ selection criteria 
measured with $K_{\mu2}$ monitor trigger data. Only statistical
uncertainties are shown. The product is $A_{K_{\mu2}}$.}
\begin{tabular}{lr}
\hline 
Cut                                     & Acceptance factor \\
\hline
Range stack track reconstruction        & $0.99993\pm0.00001$ \\
UTC-range stack track matching          & $0.99943\pm0.00002$ \\
Beam and target pattern                 & $0.15081\pm0.00018$ \\
Photon veto                             & $0.48122\pm0.00200$ \\
\hline
$A_{K_{\mu2}}$                            & $0.07253\pm0.00031$\\
\hline
\end{tabular}
\end{table}

To measure the acceptance of the range stack track reconstruction, 
a sample of $K_{\mu2}$ monitor triggers was selected by requiring 
a good track in the target and UTC, an energy deposit in the B4 hodoscope
consistent with a beam kaon and a delayed-coincidence of $>5$ ns
based on  $C_K$ and the IC. 

The acceptance associated with the matching of the range stack and UTC track 
was assessed using a $K_{\mu2}$ sample
with a good track in the range stack, a delayed-coincidence of $>5$ ns 
based on the ${\rm \check{C}}_K$ and the IC, and a single entering kaon 
selected based on the B4 energy deposit and the beam \v{C}erenkov and
wire chambers.

The acceptance factor associated with the beam and target pattern recognition
was evaluated on a sample of $K_{\mu2}$ events that were required to 
have a single entering kaon and a good track in the UTC and range stack 
with $|\cos\theta|<0.5$. In addition the momentum of the reconstructed 
track was required to be 
within two standard deviations of the expectation for 
$K_{\mu2}$ decays. There were over forty individual cuts associated
with the  beam and target  pattern recognition as described in 
Section~\ref{sec:target_cuts}. 
The majority of the individual cuts
had acceptance greater than 90\% except for the delayed-coincidence
(75.5\%) and the CCDPUL (45.1\%) requirements. 

To measure the acceptance of the standard photon veto, an additional criterion was
applied to the $K_{\mu2}$ events used for the beam and target acceptance.
As muons from $K_{\mu2}$ decay can penetrate into the barrel veto liner,
the reconstructed track was required to stop before the outermost layer
of the range stack. The acceptance factor given in Table~\ref{tab:km2acc} 
evaluated  
in this manner yielded the overall acceptance of both the online and
offline photon veto cuts as the $K_{\mu2}$ monitor trigger did not include
the photon veto.

\subsubsection{Acceptance factors from   $K_{\pi2}$ events \label{sec:acckp2}}

The $K_{\pi2}$ monitor data were used to assess the acceptance
factors associated 
with charged track reconstruction in the UTC and 
in the target. The acceptance of the veto of an additional track in the
target (OPSVETO, Section~\ref{sec:target_cuts}) was also measured with 
$K_{\pi2}$ monitors. The factors are listed in Table~\ref{tab:kp2acc} 
and described below.
\begin{table}[h]
\caption{\label{tab:kp2acc}Acceptance factors measured with $K_{\pi2}$ monitor trigger data.
Uncertainties are statistical only. The product of all factors is 
$A_{K_{\pi2}}$.}
\begin{tabular}{lr}
\hline
Cut                              & Acceptance factor \\
\hline
UTC reconstruction               & $0.94345\pm0.00019$ \\
OPSVETO                          & $0.97417\pm0.00063$ \\
Target track reconstruction      & $0.71851\pm0.00181$ \\
\hline
$A_{K_{\pi2}}$                     & $0.6604\pm0.0018$ \\
\hline
\end{tabular}
\end{table}

To measure the acceptance of the UTC reconstruction, 
events from the $K_{\pi2}$ monitor trigger were required to have
a well-reconstructed track in the range stack and agreement between
the online and offline determination of the range stack stopping counter.

For the measurement of the acceptance of the OPSVETO cut, 
in addition to the requirements described above, 
the charged track
was required to be well-reconstructed in the UTC and range stack, 
identified as a $\pi^+$ from $K_{\pi2}$ decay based on the measured range and momentum as 
well as the $\pi\to\mu\to e$ signature in the stopping counter and
kinematically consistent with the pion from a $K_{\pi2}$ decay.
Cuts were also applied to ensure a single kaon entered the target.

In addition to the requirements described above, the OPSVETO and
target photon veto cuts were applied to the $K_{\pi2}$ monitor
events to assess the cumulative acceptance of the 
ten cuts associated with target track reconstruction. These ten cuts 
were designed to reject tracks that contained an indication of 
a kink or discontinuity in the pattern of target fibers or 
target fibers with an unexpected energy deposit 
(Section~\ref{sec:target_cuts}). 
Two individual cuts with less than 90\% acceptance were the
requirement that no individual pion fiber had more than 3 MeV (89.6\%) 
and the requirement 
on the target-track fitter probability ${\cal P}(\chi^2_5+\chi^2_6+\chi^2_7)$ 
(87.4\%).

\subsubsection{Acceptance factors from   $\pi_{\rm scat}$ events}

Beam pions that scatter in the target had range, energy and
momentum spectra similar to that of pions from \KPpnn\ and were used to
determine the acceptance factors associated with the reconstruction and
identification of pions in the range stack. Table~\ref{tab:piscatacc} lists 
the acceptance factors measured using $\pi_{\rm scat}$ monitors.

Candidate events were only accepted if the pion stopped in a counter with 
an operational TD (``Good TD''). The acceptance factor associated with this
requirement was measured on a sample of $\pi_{\rm scat}$ monitor data
selected by requiring 
a single pion entering the target, 
a good outgoing track in the target, UTC and range stack,
a delayed coincidence of less than 5 ns, 
and range, energy and momentum in the signal region. 

In addition to the requirements listed above, $\pi_{\rm scat}$ monitor 
data were also required to have a good $\pi\to\mu\to e$ signature and
the pion was required to stop in a range stack counter with an operational TD 
in order to measure the acceptance factor associated with the range stack
kinematics and tracking. 
Assignment  of target fibers to the incoming and outgoing pion 
in $\pi_{\rm scat}$ was not as robust as the assignments made for kaon
decays at rest. Misassignment of target fibers yielded a larger uncertainty 
in the momentum, range and energy calculated for 
the outgoing pion in $\pi_{\rm scat}$  events. The effect of increasing or
decreasing the signal region in momentum, range and energy (Section~\ref{sec:kinematic_cuts}) 
by $\pm1$ standard deviation was used to 
estimate the systematic uncertainty in this acceptance factor.

The requirements used to assess the acceptance factor associated with a
non-operational TD in the stopping counter were supplemented by requiring 
a good track in the UTC and application of the RNGMOM cut 
in order to measure the acceptance factor associated with 
the $\pi\to\mu\to e$ signature. The cut on the measured $dE/dx$ in range
stack counters and the cuts on the consistency of the range stack and 
 UTC track were slightly correlated with the suite of cuts used to define
the $\pi\to\mu\to e$ signature~\cite{ref:tetsuro}.
The acceptance factor of the $\pi\to\mu\to e$ signature
was assessed both with and without these cuts applied to estimate the
systematic uncertainty due to these correlations. In addition, a
relative correction of $+1.4\%$ for pion decay-in-flight and pion absorption in the
stopping counter, estimated from simulation, was applied to the 
acceptance factor associated with the $\pi\to\mu\to e$ signature.
\begin{table}[h]
\caption{\label{tab:piscatacc}
Acceptance factors measured with $\pi_{\rm scat}$ monitor triggers.
The first and second uncertainties are statistical and systematic, respectively.
The assessment of systematic uncertainties is described in the text.
$A_{\pi_{\rm scat}}$ is the product of the three acceptance factors.}
\begin{tabular}{ll}
\hline
  Cut                          & \multicolumn{1}{c}{Acceptance factor} \\
\hline 
Good TD in  stopping counter   & $0.9984\pm0.0001$ \\
Range stack kinematics         & $0.8259\pm0.0013\pm0.0120$\\
$\pi\to\mu\to e$ signature     & $0.4805\pm0.0015\pm0.0160$\\
\hline
$A_{\pi_{\rm scat}}$             & $0.3980\pm0.0014\pm0.0140$ \\
\hline
\end{tabular}
\end{table}

\subsubsection{Acceptance factors from simulated events}

 Simulated \KPpnn\ events were used to evaluate 
the acceptance factors associated with the trigger, 
phase space and $\pi^+$-nuclear interactions  in the detector.
The acceptance of the $L1.1$ and $L1.2$ ($DC$) components of the trigger 
 as described in Section~\ref{sec:trigger} 
were evaluated with $K_{\pi2}$ ($K_{\mu2}$) monitors as described previously
in this Section. The acceptance of the remaining trigger components 
is given in the second row of Table~\ref{tab:umcacc}.
The phase space acceptance
includes the loss due to $\pi^+$ absorption in the stopping counter 
and decay-in-flight. Aside from $\pi^+$ absorption in the stopping counter, 
neither the trigger nor phase space acceptance include the effect of
nuclear interactions. As indicated in Section~\ref{sec:Overview}, 
the combined trigger and phase space acceptance factor 
of 11.8\% ($=0.3225\times0.3650$) 
was larger than the corresponding factor of 6.5\% for the 
\pnnone\ region~\cite{ref:pnn1PRD}. 
The acceptance factor associated with 
$\pi^+$-nuclear interactions was evaluated separately and is given
in the fourth row of Table~\ref{tab:umcacc}. For the \pnnone\ region, 
the acceptance factor due to nuclear interactions was 49.5\%~\cite{ref:pnn1PRD}.
\begin{table}[h]
\caption{\label{tab:umcacc}Acceptance factors determined from 
simulated \KPpnn\ decays. $A_{MC}$ is the product of the
three acceptance factors. The uncertainties are statistical.}
\begin{tabular}{lr}
\hline
Component                        & Acceptance factor \\
\hline
Trigger                          & $0.3225\pm0.0015$ \\
Phase space                      & $0.3650\pm0.0027$ \\
$\pi^+$-nuclear interactions     & $0.8284\pm0.0104$ \\
\hline
$A_{MC}$                          &$0.0975\pm0.0009$ \\
\hline
\end{tabular}
\end{table}

\subsubsection{Correction to the  $T\cdot2\cdot IC$ efficiency}

The $T\cdot2\cdot IC$ component of the trigger (Section~\ref{sec:trigger}) required a
coincidence between range stack counters in the same sector in the two 
innermost layers and in the IC. The simulation did not include the 
acceptance loss due to gaps between the neighboring T counters or 
due to insufficient scintillation light in the thin T counters. These 
acceptance losses were measured by using $K_{\mu2}$ and $K_{\pi2}$ decays 
in KB monitor events. The energy loss in the T counter by the charged track
differs for $K_{\mu2}$ and $K_{\pi2}$ events and simulated events were used to
obtain the average energy loss for each decay. The measured acceptance 
factors for $K_{\mu2}$ and $K_{\pi2}$ were then extrapolated to estimate 

\begin{equation}\label{eqn:t2eff}
A_{T\cdot2\cdot IC} = 0.9505\pm0.0012\pm0.0143
\end{equation}
\noindent where  a $\pm1.5\%$ systematic uncertainty was assigned to
account for the extrapolation of the UTC track to the T counter.

\subsubsection{Normalization to the   $K_{\mu2}$ branching ratio}

We assessed the fraction ($f_s$) of $K^+$ that stopped in the target by normalization
to the $K_{\mu2}$ branching ratio~\cite{ref:PDG}  as described in ~\cite{ref:pnn1PRD} 

\begin{equation} \label{eqn:fs}
f_s = 0.7740 \pm 0.0011 \ \ .
\end{equation}

\subsubsection{Confirmation of the   $K_{\pi2}$ branching ratio}

The $K_{\pi2}$ branching ratio was measured using the $K_{\pi2}$ monitor trigger 
data in order to confirm the validity of the majority of 
acceptance factors and corrections calculated 
with data and simulation. The acceptance factors associated with 
the photon veto (Table~\ref{tab:km2acc}) were not checked by this
procedure because the standard photon veto cuts 
were not applied for this measurement.
Our measurement followed the same analysis procedure as
described in ~\cite{ref:pnn1PRD} but utilized the selection criteria developed for 
the \pnntwo\  analysis. From this analysis we obtained

\begin{equation}\label{eqn:brkp2}
{\cal B}(K^+\to\pi^+\pi^0) = 0.221\pm 0.002
\end{equation}
\noindent where the uncertainty is statistical. The 6\% difference with 
the world average value~\cite{ref:PDG} of $0.209\pm0.001$ for the branching ratio
was taken into account in the assigned systematic uncertainty 
discussed in the next section.

\subsubsection{Overall acceptance and sensitivity}

The total acceptance was evaluated as the product of 
$A_{K_{\pi2}}$, $A_{K_{\mu2}}$, $A_{\pi_{\rm scat}}$, $A_{MC}$, $f_s$ and 
$A_{T\cdot2\cdot IC}$ giving  $A_{tot}=(1.37\pm0.14)\times10^{-3}$ where 
we assigned a 10\% uncertainty on the total acceptance 
to accommodate the discrepancy in ${\cal B}(K^+\to\pi^+\pi^0) $  and 
the additional systematic and statistical 
uncertainties in the acceptance evaluated in this Section. 
Based on the total exposure of $N_K={\KBLIVE}$  stopped kaons for this analysis,
the single event sensitivity (${\rm SES}\equiv 1/(N_KA_{tot})$) 
of the \pnntwo\ analysis was 
${\rm SES} = (4.28\pm0.43)\times 10^{-10}$ which can be compared 
with the SES of the 
E949 \pnnone\ analysis of $(2.55\pm0.20)\times 10^{-10}$~\cite{ref:pnn1PRD} and 
the combined SES of the previous \pnntwo\ analyses of 
$(6.87\pm0.04)\times 10^{-10}$~\cite{ref:pnn2_PLB,ref:pnn2_PRD}.

\subsection{Likelihood method}\label{sec:like}

We determined ${\cal B}(\KPpnn)$ using a likelihood method that took into account
the distributions of the predicted background and acceptance within the signal region.
The signal region was divided into nine cells with differing acceptance-to-background 
ratios as described below. The likelihood ratio $X$ was defined as

\begin{equation}\label{eqn:junk}
X 
\equiv
\prod_{i=1}^n \frac{e^{-(s_i+b_i)}(s_i+b_i)^{d_i}}{d_i!}
              \Big /
             \frac{e^{-b_i} b_i^{d_i}}{d_i !}
\end{equation}
where $s_i$ and $b_i$ were the estimated 
signal and background in the $i^{\rm th}$ cell, respectively, 
$d_i$ was the observed number of signal candidates in the $i^{\rm th}$ cell and
$n$ was the total number of cells~\cite{ref:Junk}. 
The estimated signal in each cell was given by 
$s_i \equiv  {\cal B}(\KPpnn)/{\rm SES}_i$ where
${\rm SES}_i$ was the single event sensitivity of the  $i^{\rm th}$ cell.

The division of signal region into nine cells was performed using combinations
of the decay pion kinematics (KIN), photon veto (PV), delayed coincidence (DC) 
or $\pi\to\mu\to e$ (TD)
cuts. We defined a standard ({\sl e.g.} \KINSTD) and a more restrictive or ``tight'' version of each cut 
as described previously in 
Sections~\ref{sec:kinematic_cuts} (\KIN),
 \ref{sec:pv_cuts} (\PV),
 \ref{sec:delco} (\DC) and 
 \ref{sec:muon_rej} (\TD), and 
\KINBAR\ was defined as $\KINSTD \equiv \KIN + \KINBAR$.
The signal region was defined by the application of the standard version of 
all cuts. The signal region was then subdivided into cells by the additional 
selective application
of the tight version of each cut or the inverted cut 
as shown in Table~\ref{tab:cells}. 
\begin{table} 
\caption{\label{tab:cells}The estimated signal-to-background ($s/b$) 
and background ($b$) for the nine signal region cells. The cuts defining
each cell are also given. The $s/b$ is calculated assuming 
${\cal B}({\KPpnn}) = 1.73\times 10^{-10}$. 
The uncertainties on $b$ and $s/b$ are 
omitted from the table.
In the definition column,  \KINSTD\ indicates the region defined
by the standard decay pion kinematics cut and    
\KIN\ indicates the region defined by 
the tight version of the kinematics cut.
\KINBAR\ is defined as $\KINSTD \equiv \KIN + \KINBAR$.
Analogous designations are used for the 
 photon veto (PV), delayed coincidence (DC) 
and $\pi\to\mu\to e$ (TD) cuts.}
\begin{tabular}{c|c|c|c}
\hline
Cell&   Definition                     & $b$      & $s/b$ \\
\hline
1&$\KIN\cdot\TD\cdot\DC\cdot\PV$       & 0.152    & 0.84 \\
2&$\KIN\cdot\TDBAR\cdot\DC\cdot\PV$    & 0.038    & 0.78 \\
3&$\KIN\cdot\TD\cdot\DCBAR\cdot\PV$    & 0.019    & 0.66  \\
4&$\KIN\cdot\TDBAR\cdot\DCBAR\cdot\PV$ & 0.005    & 0.57 \\
5&$\KIN\cdot\TD\cdot\DC\cdot\PVBAR$    & 0.243    & 0.47 \\
6&$\KIN\cdot\TDBAR\cdot\DC\cdot\PVBAR$ & 0.059    & 0.45 \\
7&$\KIN\cdot\TD\cdot\DCBAR\cdot\PVBAR$ & 0.027    & 0.42 \\
8&$\KIN\cdot\TDBAR\cdot\DCBAR\cdot\PVBAR$&0.007   & 0.35 \\
9&$\KINBAR\cdot\TDSTD\cdot\DCSTD\cdot\PVSTD$ & 0.379    & 0.20  \\
\hline
\end{tabular}
\end{table}

The additional signal acceptance factor and rejection (Table~\ref{tab:accrej}) 
for each of the four tight cuts 
was determined  using analogous techniques and samples 
as described in Sections~\ref{sec:acceptance} and 
\ref{sec:bkgd_evaluation}, respectively. 
Based on studies of data and simulated events,  the background 
components for which no rejection is given 
in the Table were reduced by the acceptance factor of the
particular cut. For example, the acceptance of the cell defined by 
$\KIN\cdot\TDBAR\cdot\DCBAR\cdot\PV$ \emph{relative} to the acceptance 
for the entire signal region was 
$A_{\tiny\KIN}\times (1-A_{\tiny\TD}) \times (1-A_{\tiny\DC}) \times A_{\tiny\PV}$ 
with obvious
notation, and the $K_{\pi2}$-target-scatter background component relative to the 
contribution to the entire signal region was reduced by the factor
$1/R_{\tiny\KIN} \times (1-A_{\tiny\TD}) \times (1-A_{\tiny\DC}) \times 1/R_{\tiny\PV}$.
\begin{table}[h] 
\caption{\label{tab:accrej}Additional acceptance factor ($A$) or rejection ($R$) for the 
tight version of the \KIN, \PV, \DC\ and \TD\ cuts for specific backgrounds.}
\begin{tabular}{l|llll}
\hline
                      &\multicolumn{1}{c}{\KIN}            &   \multicolumn{1}{c}{\TD}       & \multicolumn{1}{c}{\DC}       & \multicolumn{1}{c}{\PV}      \\
\hline
$A$                   & \multicolumn{1}{c}{0.812}           & \multicolumn{1}{c}{0.812}       & \multicolumn{1}{c}{0.911}     & \multicolumn{1}{c}{0.522}    \\
\hline
$R$                   & 1.63($K_{\pi2}$) & 3.08 (Muon) & 6.3 (CEX) & 2.75 ($K_{\pi2}$) \\
                      & 2.70($K_{e4}$)   &             & 1.0 (Beam)&  2.75 ($K_{\pi2\gamma}$)\\
                      & 1.20($K_{\pi2\gamma}$)&          &           & \\
\hline
\end{tabular}
\end{table}

\section{Results}
\label{sec:Results}  

In this Section, we describe
the results of examining the signal region and the evaluation 
of the \KPpnn\ branching ratio.
We also describe the evaluation of our observations within alternative models 
of {\KPpnothing}.

\subsection{Examination of the signal region}

\begin{figure}
\begin{center}
\includegraphics[width=1.1\linewidth]{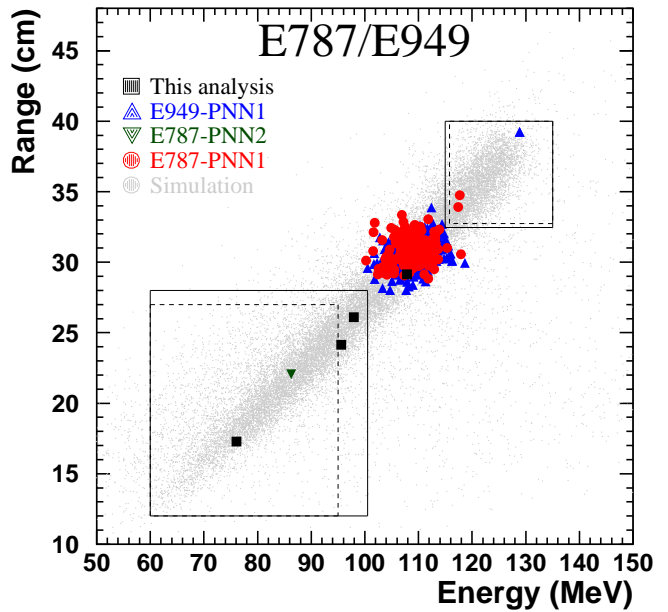}
\caption{\label{fig:rve} 
Kinetic energy {\sl vs.} range of all events passing all other 
cuts.
The squares represent the events observed by this analysis.
The circles and upward-pointing triangles represent the events 
observed by the E787 and E949 \pnnone\ analyses, respectively.
The downward-pointing triangle represent the 
events observed by the E787 \pnntwo\ analyses. 
 The solid (dashed) lines 
represent the limits of the \pnnone\ and \pnntwo\ signal regions for 
the E949 (E787) analyses. Despite the smaller signal region 
in  $R_\pi$ {\sl vs.} $E_\pi$, 
the \pnnone\ analyses were 4.2 times more sensitive than the \pnntwo\ analyses. 
The points near $E_\pi = 108$ MeV were \KPtwo\ decays that survived the photon 
veto cuts
and were predominantly from the \pnnone\ analyses due to the higher 
sensitivity and the less stringent photon veto cuts.  
The light gray points are simulated \KPpnn\ events that were
accepted by the \oneortwo\ trigger.
}
\end{center}
\end{figure}

After completion of the background and acceptance analyses, all selection criteria
were applied to the \oneortwo\ trigger data. Three signal candidate events were 
observed.
 Some measured  properties of the three events are listed in Table~\ref{tab:sigkin}.
\begin{table}[h] 
\caption{\label{tab:sigkin}
The cell (Table~\ref{tab:cells}) and the 
momentum, range and energy of the $\pi^+$ in the three signal candidate
events A, B and C.
The measured decay times are in units of $K^+$, $\pi^+$ and $\mu^+$ 
lifetimes~\cite{ref:PDG}, appropriately.}
\begin{tabular}{l|rrr}
\hline
  Event             &\multicolumn{1}{c}{A} &\multicolumn{1}{c}{B} &\multicolumn{1}{c}{C} \\
\hline
Cell               & \multicolumn{1}{c}{9}   & \multicolumn{1}{c}{5}  & \multicolumn{1}{c}{7} \\
\hline
Momentum (MeV$/c$)      & 161.5                & 188.4                &  191.3  \\
Range (cm)              &  17.3                &  24.2                &   26.1 \\
Kinetic energy (MeV)    &  76.1                &  95.6                &   97.9  \\
\hline
$K^+$ decay time          &   0.30                &  1.27                &    0.42  \\
$\pi^+$ decay time        &   0.86                &  0.64                &    0.39  \\
$\mu^+$ decay time        &   2.71                & 1.03                 & 4.33  \\
\hline
\end{tabular}
\end{table}
The plot of the kinetic energy {\sl vs.} range of the three events along
with the events found in the previous {\pnnone}~\cite{ref:pnn1PRD} and
{\pnntwo}~\cite{ref:pnn2_PLB,ref:pnn2_PRD} analyses are shown in 
Figure~\ref{fig:rve}.
The candidates' measured properties used in the selection criteria 
were consistent with  the 
expected distributions for \KPpnn\ decay signal. 
There was no observed activity in the kaon fibers at 
the time of the $\pi^+$ for any of the three candidates according to 
the CCDPUL analysis. 
All three events failed one or more of the tight cuts
described in Section~\ref{sec:like}. 
The $\pi^+$ momentum of event A failed the tight momentum cut of 165 MeV$/c$. 
Events B and C failed the tight photon veto cut due to energy deposits of 2.4 
and 2.1 MeV in the EC 
above the 1.7 MeV threshold, respectively, 
and events A and C failed the tight \DC\ cut of 6 ns on the kaon decay time.

\subsection{The \boldmath \KPpnn\  branching ratio}

The central value of the \KPpnn\ branching ratio was taken to be 
the value of ${\cal B}(\KPpnn)$ that maximized 
the likelihood ratio 
$X$ (Equation~(\ref{eqn:junk})). 
For the three events observed by this analysis, 
we determined ${\cal B}(\KPpnn) = \BRthis$ where the quoted 68\% confidence level 
interval was determined from the behavior of $X$ as described in \cite{ref:Junk} 
and took into account both the statistical and systematic uncertainties. 
The systematic uncertainties included the 10\% uncertainty in the acceptance
as well as the uncertainties in the estimation of the background components.
The inclusion of systematic uncertainties had a negligible effect on 
the confidence level interval due to the poor statistical precision
inherent in a three event sample. The probability that these three events
were due to background only, given the estimated background in each cell 
(Table~\ref{tab:cells}), was 0.037.

When the results of the previous \pnnone\ and \pnntwo\ 
analyses~\cite{ref:pnn1PRD,ref:pnn2_PLB,ref:pnn2_PRD} were combined with 
the results of this analysis, 
we determined ${\cal B}(\KPpnn) = \BRall$~\cite{ref:web949}.
Systematic uncertainties were treated as described above when 
performing the combination, except that we assumed a correlated 10\% 
uncertainty for the acceptance assessed by each analysis. 
The probability
that all seven events were due to background only 
(background and SM signal) 
was estimated to be 0.001 (0.073).

The \KPpnn\  branching ratio for each of two pion momentum regions 
are
given in Table~\ref{tab:DR} along with the SM predictions. 
The combined E787 and E949 \pnntwo\ (\pnnone) data were 
used for the branching ratio in the  $[130,205]$ ($[205,227]$) MeV$/c$ region.
The limits of the two momentum regions were determined by the requirements on 
the reconstructed pion region, the detector resolution and the desire to have
contiguous, non-overlapping regions. 
The boundary between the two regions of 205 MeV/$c$ was determined by 
the lower and upper limits on the reconstructed 
$\pi^+$ momentum that were set to be approximately 2.5 standard
deviations from the nominal $K_{\pi2}$ momentum for 
the \pnnone\ and \pnntwo\ analyses, respectively.
The lower limit of 130 MeV$/c$ was determined by the \pnntwo\ requirement
on the reconstructed pion momentum. The upper limit of 227 MeV$/c$ is the kinematic
limit for the \KPpnn\ decay.
\begin{table}[h]
\caption{\label{tab:DR}The measured 
 branching ratios  in units of  $10^{-10}$ assuming
a standard model, scalar or tensor interaction~\cite{ref:web949} in two momentum regions as
described in the text.
The SM prediction was taken from \cite{ref:TH2} and scaled to the two 
pion momentum regions using the SM spectral shape for 
{\KPpnn}.
The 68\% and 90\% CL intervals for the double ratio, described in the text,
are also given.}
\begin{tabular}{c|r|r|r|r}
\hline
Momentum  &       \multicolumn{1}{c|}{SM} & \multicolumn{3}{c}{Interaction}\\
range (MeV$/c$) & \multicolumn{1}{c|}{prediction}& \multicolumn{1}{c}{SM} & \multicolumn{1}{c}{Scalar} & \multicolumn{1}{c}{Tensor}   \\
\hline
$[130,205]$    &$0.49\pm0.04$  &   $2.91^{+4.02}_{-1.79}$      &$0.59^{+0.59}_{-0.35}$  & $0.49^{+0.45}_{-0.31}$ \\[1mm] 
$[205,227]$    &$0.28\pm0.02$  &  $0.49^{+0.45}_{-0.29}$       &$3.50^{+4.68}_{-2.10}$ & $3.10^{+4.23}_{-1.87}$  \\[1mm] 
\hline
\multicolumn{1}{r}{Double ratio}& {68\% CL}               &(0.56,6.15)     & (0.29,3.32)  & (0.08,0.88) \\
\multicolumn{1}{l}{}         & {90\% CL}               &(0.24,17.6)     & (0.13,9.55)  & (0.04,2.66) \\
\hline
\end{tabular}
\end{table}

The 90\% CL upper limit ${\cal B}(\KPpnn) < \UCLall$ was also determined
and  can be used to 
calculate a model-independent upper limit of \UCLgn\ on the $CP$-violating process 
\KZpnn\cite{Grossman:1997sk}. This limit is substantially smaller than
the current experimental limit of 
${\cal B}(\KZpnn) < 670 \times 10^{-10}$~\cite{Ahn:2007cd}.

\subsection{Alternative model interpretations}

\begin{figure}[h]
  \includegraphics[width=\linewidth]{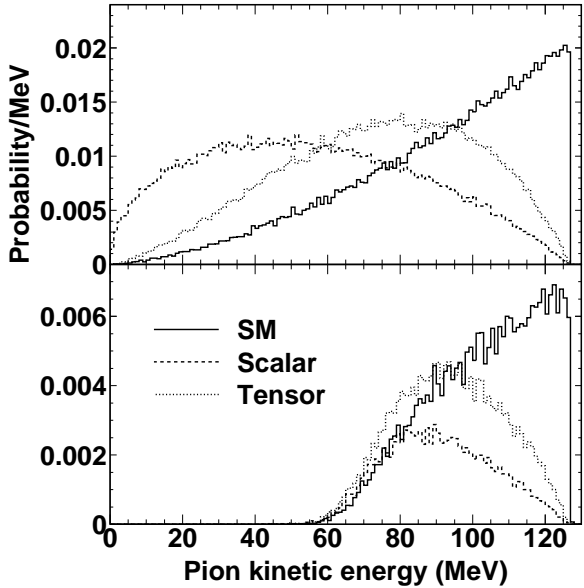}
  \caption{\label{fig:stx} (Top) The generated $\pi^+$ kinetic energy  spectrum 
    for SM (solid), scalar (dashed) and tensor (dotted) interactions.
    (Bottom) The distributions for events passing the \oneortwo\ trigger.
    }
\end{figure}

The combined results of the E787 and E949 experiments can also be interpreted
assuming a scalar or tensor interaction. The spectra are
compared with the SM spectrum in Figure~\ref{fig:stx}.
Using the same treatment of the data
as for the standard model interpretation, the branching ratio
for the scalar and tensor spectra alone were 
${\cal B}_{\rm sc}(\KPpnn) = \BRscalar$ and
${\cal B}_{\rm te}(\KPpnn) = \BRtensor$, respectively, 
or 
${\cal B}_{\rm sc}(\KPpnn) < \UCLscalar$ and 
${\cal B}_{\rm te}(\KPpnn) < \UCLtensor$ at 90\% CL.

The consistency of the distribution of the observed events 
with the shape of the SM, scalar or tensor spectrum was evaluated using
the double ratio 
\begin{equation}\label{eqn:DR}
{\rm DR}
\equiv 
\frac{{\cal B}(130,205)}{{\cal F}(130,205)}
\Big /
\frac{{\cal B}(205,227)}{{\cal F}(205,227)}
\end{equation}
where ${\cal B}(P_1,P_2)$ is the  branching ratio 
and ${\cal F}(P_1,P_2)$ is the fraction of phase space 
in the momentum range $(P_1,P_2)\ {\rm MeV}/c$. 
The ratio ${\cal F}(205,227)/{\cal F}(130,250)$ 
is 0.084, 0.014 and 0.59 for the scalar, tensor 
and standard model spectra, respectively. 
If the distribution
of observed events were consistent with the shape of the 
assumed spectrum, then the double ratio would be unity. 
The 68\% and 90\% CL intervals of the DR and the 
 branching ratios  are given in Table~\ref{tab:DR}.

\begin{figure}[h]
  \includegraphics[width=\linewidth]{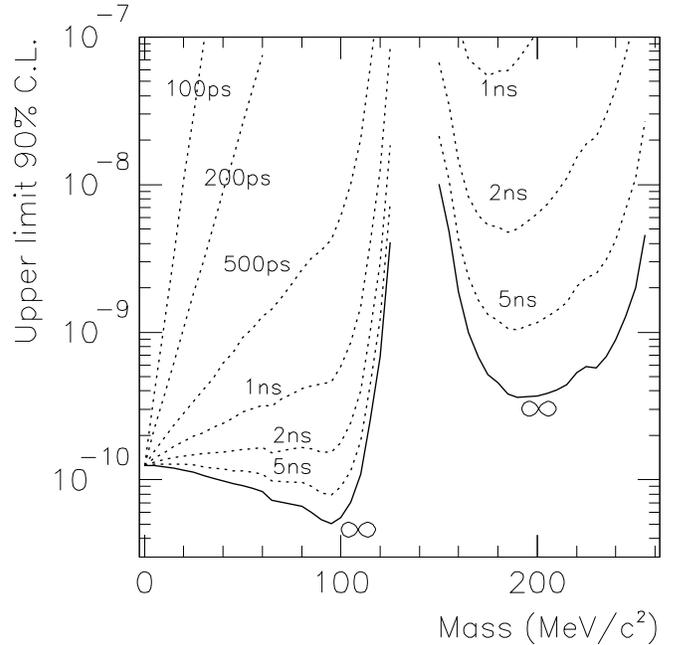}
  \caption{\label{fig:kpx}The solid line represents the 90\% CL upper limit
    on ${\cal B}(K^+\to\pi^+X)$ as a function of the mass of $X$ 
    assuming $X$ is stable.
  The dashed lines represent the 90\% CL for $X$ with the indicated lifetimes 
as described in the text.}
\end{figure}

The data have also been interpreted in the two-body decay model, 
$K^+\to\pi^+ X$, where $X$ is a massive non-interacting particle, either
stable or unstable. 
The 90\% CL upper limit on the branching ratio 
as a function of the mass of $X$ is shown
in Figure~\ref{fig:kpx}. For the case of an unstable $X$, the decay of $X$
was assumed to be detected and vetoed with 100\% efficiency if $X$ decayed
within the outer radius of the BV. 
The E949 limit of ${\cal B}(\pi^0\to\nu\bar\nu) < 2.7\times 10^{-7}$ 
at 90\% CL~\cite{ref:pi0nn} can be combined with the 
world average value of ${\cal B}(K^+\to\pi^+\pi^0)$~\cite{ref:PDG} to set a 90\% CL limit
of ${\cal B}(K^+\to\pi^+ X) < 5.6\times10^{-8}$ for $M_X = M_{\pi^0}$ with $X$ stable.

The limit in the Figure~\ref{fig:kpx} can also be interpreted as a 
limit on the product of branching ratios 
${\cal B}(K^+\to\pi^+ P^0)\times{\cal B}(P^0\to\nu\bar\nu)$ for a 
hypothetical, short-lived particle $P^0$. 
The HyperCP collaboration observed three events consistent
with $\Sigma^+\to pP^0$ with $P^0\to\mu^+\mu^-$  having a mass 
$M(P^0) = 214.3\pm0.5\ {\rm MeV}/c^2$~\cite{Park:2005eka}.
A mass of $214.3\ {\rm MeV}/c^2$ would correspond to a 
recoiling $\pi^+$ momentum, range and energy of 
$170.1\ {\rm MeV}/c$, 19.5 cm and $80.5\ {\rm MeV}$, respectively,
  in a two-body $K^+$ decay.
Of the four events in the \pnntwo\ region observed in
E787~\cite{ref:pnn2_PLB} and E949, the closest was 
the present candidate A that differed from the expected $\pi^+$ momentum, 
range and energy
by 3.7, 2.4 and 1.5 standard deviations, respectively.
\section{Conclusion}
\label{sec:Conclusion}  

\subsection{Summary}\label{sec:summary}

The branching ratio  of the decay \KPpnn\ is precisely predicted in the 
standard model to be ${\cal B}(\KPpnn) = \BRSM$~\cite{ref:TH2}. As the decay 
is a flavor-changing neutral current process and thus 
sensitive to new physics effects~\cite{Buras:2004uu}, it
represents an opportunity to unambiguously probe for 
physics beyond the standard model.

The E787 and E949 experimental program at Brookhaven National Laboratory
has demonstrated the feasibility of observation of the rare decay 
\KPpnn\ using  stopped kaons  despite the 
challenging  experimental signature. 
The analysis presented in this article has established that backgrounds can be
reduced to a reasonable level while maintaining signal acceptance 
to enable the \pnntwo\ region to be 
a viable supplement to the \pnnone\ region~\cite{ref:pnn1PRD} in 
the measurement of the \KPpnn\ branching ratio. 
The branching ratio ${\cal B}(\KPpnn)=\BRall$ for the seven events observed
by E787 and E949 is consistent with the SM  expectation.

\subsection{Final comments}\label{sec:lessons}

We briefly summarize some  important issues that emerged
from the E787 and E949 searches for \KPpnn\ and the measurement
of ${\cal B}(\KPpnn)$ in a stopped $K^+$ experiment:
\begin{enumerate}
  \item Importance of blind analysis. The analysis procedure of 
    concealing or obscuring
    the contents of the signal region~\cite{ref:pnn1_2,Arisaka:1992xy} 
    are now well-established and widespread in 
    particle physics. The main benefit is to 
    avoid or minimize bias
    in the selection criteria. This is particularly important for a rare 
    process such
    as \KPpnn\ with a poor experimental signature that requires many 
    individual cuts. 

  \item Use of data to estimate background and acceptance. 
    In conjunction with the blinding of the signal region, the division of the
    data into 1/3 and 2/3 samples and the use of two powerful independent cuts 
    provides background estimates that take into account instrumental effects 
    not present in simulation and provides sensitivity to background beyond 
    that of the \KPpnn\ sample as first noted in \cite{ref:pnn1_2}. 
    The bifurcation technique also permits examination of events that occur
    near the signal region to validate the background estimates 
    and investigate unforeseen sources of background.

  \item Unforeseen acceptance losses. In the E787 proposal~\cite{ref:e787prop}, 
    only the \pnnone\ analysis
    region was considered to be viable 
    and the estimated acceptance was 1.5\%. This can be 
    compared to the E949 \pnnone\ acceptance of 0.22\%~\cite{ref:pnn1PRD} and 
    0.14\% for the present analysis (Section~\ref{sec:acceptance}). 
    Accidental activity reduced the acceptance factors associated with 
    vetoing. Acceptance was also reduced
    due to the cuts to suppress background,
    particularly related to muons in the \pnnone\ region and 
    pion scattering in the \pnntwo\ region, 
    that needed to be more stringent
    than anticipated as well as due to cuts to suppress beam-related background
    that were not considered in the proposal.
    
  \item Importance of redundancy 
    in detector systems. 
    In E949 and its predecessor E787, since nearly every detector element 
    participated in the veto function for dealing with 
     additional particles, 
     the loss of any element represented a reduction 
    in background rejection and motivated the redundant use of ADCs, TDCs and 
    high-speed waveform digitizers (CCDs or TDs).

  \item Need for 4-$\pi$ sr photon veto coverage.
    The early \KPpnn\ counter experiments showed the benefits of photon veto
    capability over the full 4-$\pi$ sr solid angle~\cite{ref:cable,ref:pang}.
    Each modification or upgrade of the original E787 experiment included
    photon veto enhancements.
    For example, E949 contained upgrades with respect to E787 that sought to 
    further suppress the contribution of the 
    $K_{\pi2}$ background to both the \pnnone\ and \pnntwo\ analyses.
    The barrel veto liner (BVL) improved the photon veto rejection in the barrel 
    region by a factor of 2 and the active degrader (AD) increased the 
    photon veto rejection for the \pnntwo\ analysis by 2  (Section~\ref{sec:pv_cuts}).
    
\end{enumerate}

Extrapolating from the E787 and E949 experience, it would be 
possible to extend the stopped $K^+$ experiment approach to 
yield orders of magnitude more observed \KPpnn\  decays~\cite{ref:goldenbook}. 
As indicated above, the high rate of kaon interactions at 710 MeV$/c$ 
resulted in reduced acceptance due to accidental spoiling of good 
events and increased backgrounds requiring highly  restrictive cuts. 
If a higher  primary proton beam intensity were available (with a high 
duty factor),  lower momentum kaons, {\sl e.g.} 450 MeV$/c$, would result 
in a much higher stopping fraction and  reduced interactions which 
cause accidentals.  To take advantage of this effect, a shorter 
particle-separated beam line ({\sl e.g.} 13 m compared to the present 19 m), 
would be necessary to reduce attenuation of the kaon beam due to decay.   
Other improvements which would result in increased acceptance might involve 
use of a higher magnetic field (1T $\rightarrow$ 3T) allowing  a more compact 
detector with  improved  momentum resolution, finer segmentation of the range 
stack, a more hermetic, thicker fully active veto detector ({\sl e.g.} crystals or 
liquid xenon), and an improved target.

\begin{acknowledgments} 
We gratefully acknowledge the support and efforts of  
the BNL Collider-Accelerator Department
for the high quality $K^+$ beam delivered. 
We also recognize the substantial contributions made by 
the participants of  E787 without which this work would 
not have been feasible, as well as the excellent technical 
and engineering support provided by all collaborating institutions
including
P. Bichoneau, 
R. Bula,      
M. Burke,        
M. Constable,       
H. Coombes,       
J. Cracco,  
A. Daviel,        
H. Diaz,  
C. Donahue,  
E. Garber, 
C. Lim,      
A. Mango, 
G. Munoz,  
H. Ratzke, 
H. Sauter,  
W. Smith,  
E. Stein   
and
A. Stillman,  
This research was supported in part by the U.S.  
Department of Energy, the Ministry of Education,  
Culture, Sports, Science and Technology of Japan through  
the Japan-U.S. Cooperative Research Program in High Energy  
Physics and under Grant-in-Aids for Scientific Research,  
the Natural Sciences and Engineering Research Council and  
the National Research Council of Canada, the Russian 
Federation State Scientific Center Institute for High Energy  
Physics, and the Ministry of Science and Education of the Russian Federation. 
S. Chen was also supported by Program for New Century Excellent 
Talents in University from the Chinese Ministry of Education. 
\end{acknowledgments}

\bibliography{main_pnn2}

\begin{thebibliography}{64}
\bibitem[{ref({\natexlab{a}})}]{ref:footnote}
\bibinfo{note}{This method of assigning systematic uncertainty was intended to
  define a range that included the actual value of the background.}
\end{thebibliography}

\end{document}